\documentclass[11pt]{article}

\usepackage[letterpaper,margin=1in,headheight=16pt,headsep=16pt]{geometry}
\usepackage[T1]{fontenc}
\usepackage[utf8]{inputenc}
\usepackage{amsmath}
\usepackage{subcaption}
\usepackage{times}
\usepackage{newtxmath}
\usepackage{microtype,graphicx,booktabs,tabularx}
\usepackage{xcolor,tcolorbox}
\usepackage{enumitem}
\usepackage{fancyhdr}
\usepackage{caption}
\usepackage{multirow}
\tcbuselibrary{breakable,skins}
\usepackage{xcolor}
\definecolor{linkblue}{HTML}{1780FC}
\usepackage[colorlinks=true, urlcolor=linkblue, citecolor=black]{hyperref}

\definecolor{updateyellow}{RGB}{255,247,173}

\newcommand{\reporttitle}{AsterMind V1: Exploring Space Intelligence}
\newcommand{\reportlogofile}{Figures/AsterMind-logo.png}

\newcommand{\reportlogo}[1]{%
  \begingroup
  \setlength{\fboxsep}{0pt}%
  \setlength{\fboxrule}{0.3pt}%
  \raisebox{-0.12\height}{%
    \IfFileExists{\reportlogofile}{%
      \includegraphics[height=#1]{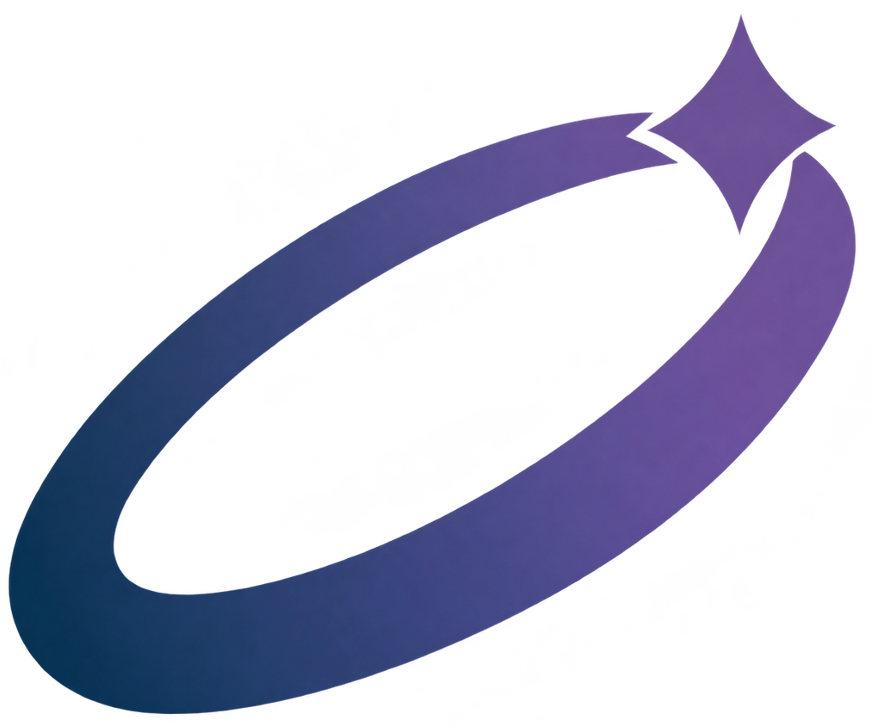}%
    }{%
      \fbox{\rule{0pt}{#1}\rule{#1}{0pt}}%
    }%
  }%
  \endgroup
}

\newcommand{\reportpagestyle}{%
  \fancyhf{}%
  \fancyhead[C]{%
    \normalfont\fontsize{10}{12}\selectfont
    \reportlogo{9pt}\hspace{0.35em}\reporttitle
  }%
  \fancyhead[R]{%
    \normalfont\fontsize{9}{11}\selectfont\scshape Technical Report
  }%
  \fancyfoot[C]{\normalfont\small\thepage}%
  \renewcommand{\headrulewidth}{0.4pt}%
  \renewcommand{\footrulewidth}{0pt}%
}
\reportpagestyle
\fancypagestyle{plain}{\reportpagestyle}

\fancypagestyle{firstpage}{%
  \fancyhf{}%
  \fancyfoot[C]{\normalfont\small\thepage}%
  \renewcommand{\headrulewidth}{0pt}%
  \renewcommand{\footrulewidth}{0pt}%
}

\newsavebox{\reporttitlebox}
\makeatletter
\newcommand{\reportsubtitle}[1]{\gdef\@reportsubtitle{#1}}
\newcommand{\@reportsubtitle}{}

\renewcommand{\maketitle}{%
  \par\begingroup
  \centering
  \setlength{\parskip}{0pt}%

  \hrule height 2pt
  \vspace{16pt}

  \sbox{\reporttitlebox}{%
    {\normalfont\fontsize{17.28}{21}\selectfont\mdseries\scshape
     \reportlogo{20pt}\hspace{0.35em}\@title}%
  }%
  \ifdim\wd\reporttitlebox>0.95\linewidth
    \resizebox{0.95\linewidth}{!}{\usebox{\reporttitlebox}}%
  \else
    \usebox{\reporttitlebox}%
  \fi
  \par
  \vspace{16pt}

  \hrule height 2pt
  \vspace{12pt}

  {\normalfont\fontsize{10}{12}\selectfont\mdseries\scshape
   \@reportsubtitle\par}
  \vspace{20pt}

  {\normalfont\fontsize{10}{12}\selectfont\bfseries
   \@author\par}
  \vspace{18pt}
  \endgroup
}
\makeatother

\renewenvironment{abstract}
{%
  \small
  \begin{center}
    \bfseries\abstractname
  \end{center}
  \begin{list}{}{%
    \setlength{\rightmargin}{\leftmargin}
    \setlength{\itemindent}{0pt}
    \setlength{\listparindent}{0pt}
    \setlength{\parsep}{5pt}
  }
  \item\relax
}
{%
  \end{list}
}

\makeatletter
\renewcommand{\@makefntext}[1]{%
  \noindent\hspace*{0.8em}\@makefnmark\hspace{0em}#1%
}
\makeatother

\renewcommand{\footnoterule}{%
  \kern-3pt
  \hrule width \columnwidth height 0.4pt
  \kern 2.6pt
}

\usepackage{calc}
\newcommand{\commcolhead}[1]{%
  \makebox[\maxof{\widthof{#1}}{\widthof{12020.07}}][c]{#1}}

\title{\reporttitle}
\reportsubtitle{Technical Report on AsterMind V1}

\usepackage{marvosym}

\author{%
  {\bfseries
    Lidong Guo\textsuperscript{1*},
    Yuze Liu\textsuperscript{1,3*},
    Jinwen Huang\textsuperscript{1,2*},
    Teng Jia\textsuperscript{1,4*},
    Qian Wang\textsuperscript{1,5*}\\[2pt]
    Shengen Yan\textsuperscript{2},
    Shengqing Shi\textsuperscript{1},
    Yi Shan\textsuperscript{1},
    Zhenhua Zhu\textsuperscript{2\,\Letter},
    Yu Wang\textsuperscript{2}%
  }\\[6pt]
  {\normalfont\small
    \textsuperscript{1}GUANYU,
    \textsuperscript{2}Tsinghua University,
    \textsuperscript{3}University of Chinese Academy of Sciences\\[2pt]
    \textsuperscript{4}University of Electronic Science and Technology of China,
    \textsuperscript{5}Beijing University of Posts and Telecommunications
  }\\[4pt]
  {\normalfont\footnotesize
    \textsuperscript{*}These authors contributed equally\quad
    \textsuperscript{\Letter}Corresponding author%
  }%
}

\date{}
\usepackage{float}
\usepackage{booktabs}
\usepackage{multirow}
\usepackage{graphicx}
\begin{document}

\maketitle
\thispagestyle{firstpage}

\begin{abstract}
Rapidly growing demand for large language model (LLM) inference is increasing the power and infrastructure required to scale serving capacity. Orbital computing offers a new deployment environment in low Earth orbit (LEO), where onboard solar generation can reduce dependence on terrestrial power and datacenter infrastructure. However, provisioning compute and power in orbit does not guarantee usable LLM serving capacity for terrestrial users. Orbital motion continuously changes ground access and inter-satellite communication paths, while request execution changes compute load and reusable session state. Quantifying this gap and identifying the underlying bottlenecks require these coupled dynamics to be modeled jointly.

We introduce AsterMind, an extensible constellation-scale simulation framework that jointly models orbital dynamics, computation and communication to evaluate end-to-end orbital LLM serving. AsterMind constructs a time-resolved hierarchical constellation model which is organized into three levels: satellite, formation, and constellation. It also incorporates a state-coupled runtime simulation engine to replay geo-temporal workloads and simulate end-to-end request execution on a shared event timeline. Using AsterMind, users can configure simulation scenarios, integrate custom strategies, and compare alternative designs under consistent workloads and constraints. These capabilities support researchers and engineers in carrying out work including communication system verification, computing payload design and constellation planning. AsterMind is open-sourced at \href{https://github.com/GUANYU-Tech/AsterMind}{https://github.com/GUANYU-Tech/AsterMind}.


\end{abstract}

\section{Introduction}
\label{sec:introduction}

Rapidly growing demand for large language model (LLM) inference is imposing increasing demand on the power supply and physical infrastructure required to scale LLM serving capacity. Orbital computing offers a new avenue for expanding artificial intelligence (AI) infrastructure through onboard solar generation, reducing dependence on terrestrial power and datacenter construction. This opportunity has attracted growing industry interest, exemplified by Google's Project Suncatcher and SpaceX's plans for AI satellites~\cite{spacex_ai1,google_suncatcher}.

Building on this industry momentum, we envision an orbital LLM serving system deployed in low Earth orbit (LEO). A compute constellation is organized into satellite formations, referred to as compute cells, each comprising one or more accelerator-equipped satellites. Requests from terrestrial users are forwarded through ground gateways and relayed by communication satellites to one compute cell. The compute cell then executes LLM inference and returns the response to the user. Three developments support the feasibility of such an orbital LLM serving system: solar energy availability, improving launch economics, and advances in onboard computing. Firstly, sunlight exposure in LEO can be 4--5$\times$ longer than that on the ground, together with 36\% higher photovoltaic condition~\cite{nasa_solar_irradiance,astm_g173_2023}. Secondly, reusable launch vehicles and standardized rideshare services can reduce the launch cost. With first-stage recovery and reuse, analyses estimate recurring launch costs of below \$1,000 per kilogram~\cite{etzenbach2025reuse}. Thirdly, for the radiation-induced faults, a carefully selected commercial off-the-shelf (COTS) platform was validated in a 32-hour heavy-ion irradiation experiment~\cite{bruhn2020radiation}.

Prior work has established technical foundations for LLM inference in space. However, an operational LLM serving system must coordinate communication and compute resources to meet latency and service continuity requirements. Even under the same constellation layout and hardware configuration, requests arriving at different times can encounter different access and execution conditions.

\begin{figure}[t]
  \centering
  \includegraphics[width=\linewidth]{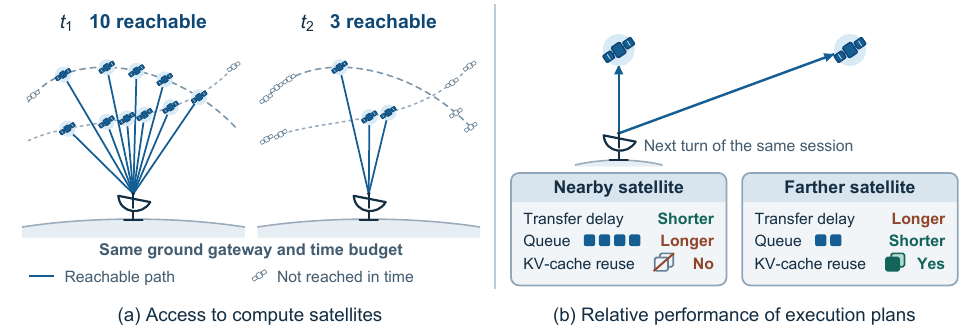}
  \caption{Time-varying access to compute satellites and execution trade-offs. (a) The number of satellites reachable within a fixed time budget changes over time. (b) Shorter queues and KV-cache reuse can offset longer transfer delays.}
  \label{fig:motivation}
\end{figure}

\textbf{Access to compute satellites.}
The set of compute satellites reachable within a given time budget depends on both the waiting time for ground access and the forwarding delay. Direct ground access is limited to visibility windows. For a satellite in a 550-km orbit with a period of approximately 95.5 minutes, an overhead pass provides only about 4.5 minutes of visibility. For inter-satellite forwarding, relative motion between compute and communication satellites changes the available relay paths and link distances, affecting propagation delay and achievable data rates. In Figure~\ref{fig:motivation}(a), two requests arriving at the same ground gateway at $t_1$ and $t_2$ can reach ten and three compute satellites, respectively, within the same time budget.

\textbf{Relative performance of execution plans.}
When several compute satellites are reachable, the response latency for a request depends on transfer delays, queueing times, and the LLM inference latency. Orbital motion changes the available communication paths and their transfer delays. At runtime, compute load affects queueing time, while the length of the reusable KV-cache prefix affects how much computation is required. The relative performance of execution plans can therefore change as orbital and runtime states change. As shown in Figure~\ref{fig:motivation}(b), the nearby satellite can be reached sooner but has a longer queue and no reusable KV-cache prefix. The farther satellite can complete the request sooner if time saved through shorter queueing and KV-cache reuse exceeds the additional communication delay.

The above effects raise a fundamental question. \textit{How can we systematically explore the LLM serving capacity supported by orbital compute and communication resources across different configurations and policies?} These effects are difficult to assess experimentally before costly in-orbit deployment. Physical payload hardware is difficult to replace after launch, while repeated constellation reconfiguration requires maneuvering time. A terrestrial prototype with a fixed network topology cannot directly reproduce the time-varying ground access and inter-satellite connectivity in orbit. Comparing deployment configurations and runtime policies therefore requires end-to-end simulation that jointly models orbital dynamics, computation, and communication as each request proceeds through ground access, forwarding, queueing, inference, and response delivery. Existing simulation tools support orbital analysis~\cite{stk}, network simulation~\cite{hypatia,starperf}, and onboard-system modeling~\cite{oec}. However, used in isolation, these tools leave the interaction between evolving orbital conditions and LLM inference progress unmodeled.

We introduce AsterMind, an extensible constellation-scale simulation framework that jointly models orbital dynamics, computation and communication to evaluate end-to-end orbital LLM serving. Its simulation backend comprises two components:

\textbf{(1) Offline Constellation Builder.} The builder constructs a hierarchical constellation model at the satellite, formation, and constellation levels from scenario configurations. It precomputes time-resolved orbital geometry and candidate communication links over the simulation horizon.

\textbf{(2) State-Coupled Runtime Simulation Engine.} The runtime engine replays geo-temporal workloads on the generated constellation model under configurable deployment, routing and scheduling policies. It simulates end-to-end request execution on a shared event timeline, coupling communication and inference events with updates to resource occupancy and session state.

AsterMind V1 provides three key features for exploring space intelligence:

\begin{itemize}[leftmargin=*, itemsep=2pt, topsep=3pt]
  \item \textbf{End-to-end simulation across deployment scales.} AsterMind supports single-satellite, satellite formation, and constellation-scale scenarios within a common simulation framework. It supports the simulation of end-to-end request execution on a shared event timeline.

  \item \textbf{Multi-dimensional state modeling.} Building on the scenario generated 
  by the offline constellation builder, AsterMind jointly models orbital dynamics, computation, and communication. It captures how the time-varying connectivity, bandwidth contention, compute load, and session-state reuse interact throughout request execution. 

  \item \textbf{Configurable scenarios and extensible policies.} Users can construct scenarios by specifying constellation layouts, hardware resources and workloads. They can also define and integrate custom deployment, routing, and scheduling policies to explore new strategies and compare alternative designs.
\end{itemize}



\section{Background and Motivation}
\label{sec:background}

\subsection{LLM Serving}
\label{sec:background-llm}

LLM inference consists of two stages: Prefill processes the input prompt, while the decode stage generates output tokens one at a time. The key-value (KV) cache stores attention key and value tensors for reuse during generation. A session is a conversation containing one or more turns. Requests with a shared prompt prefix can reuse cached KV states to reduce prefill computation, both within a multi-turn session and across sessions. LLM requests come from geographically distributed users, with volumes varying across regions and throughout the day~\cite{anthropic2026economicindex,zou2026officehours}. Requests also differ in prompt length, output length, and reusable context, so equal request counts may exhibit different computational demands.

Serving performance is evaluated by response latency and the rate of completed requests meeting the SLO. Time-to-first-token (TTFT) measures the delay from request submission to receipt of the first output token, while end-to-end (E2E) latency measures the delay to receipt of the complete response. For a response containing $N > 1$ output tokens, the average time-per-output-token (TPOT) is $(T_{\mathrm{E2E}}-T_{\mathrm{TTFT}})/(N-1)$~\cite{nvidia_nim_metrics}. An SLO specifies performance requirements, such as limits on TTFT and TPOT~\cite{distserve}. SLO goodput is defined as the number of completed requests per second meeting these limits~\cite{wang2024revisiting}. To reduce response latency and improve serving throughput, distributed LLM serving systems consider both replica load and prefix reuse when assigning requests to model replicas~\cite{srivatsa2025preble,xia2025skylb}.

\subsection{Orbital Computing Infrastructure}
\label{sec:background-orbit}

Orbital computing uses processors aboard satellites to execute computing workloads. A compute satellite consists of a computing payload and a spacecraft bus. The payload includes processors and storage, while the bus provides power, thermal control, communications, and orbit control~\cite{esa_gaia_service_module,nasa_smallsat_avionics}. Starcloud-1 carried an NVIDIA H100 GPU into space in November 2025~\cite{starcloud1_2025}. Such hardware introduces substantial power and thermal loads. For reference, the H100 SXM has a maximum thermal design power (TDP) of 700~W~\cite{nvidia_h100}, while an NVL72 rack is rated at 140~kW, illustrating the wide range of power requirements across hardware configurations~\cite{supermicro_gb300_nvl72}. Supporting these loads requires power generation, energy storage, heat transport, and radiators~\cite{architecting_space_microdatacenters,orbital_datacenters}.

Solar arrays supply operating power and recharge batteries during sunlight, while batteries support operation during eclipse. Orbital configuration affects the sunlight exposure and eclipse duration, and therefore influences power-system sizing~\cite{esa_power_systems,nasa_smallsat_power}. The surrounding vacuum precludes convective heat transfer to the external environment, so spacecraft must transport heat to radiating surfaces and reject it into space~\cite{esa_thermal_control,nasa_smallsat_thermal}. S$\mu$DC estimates that approximately 4~m$^2$ of radiator panel area can reject 4~kW of heat at 45$^{\circ}$C, assuming an emissivity of 0.86 and both panel faces viewing deep space~\cite{architecting_space_microdatacenters}. Increasing the computing load requires larger solar arrays, batteries, and radiators, increasing spacecraft mass and required launch capacity~\cite{architecting_space_microdatacenters,orbital_datacenters}.

Ground-satellite links (GSLs) connect ground gateways to satellites, commonly using radio-frequency communication. Inter-satellite links (ISLs) relay data between satellites using lasers~\cite{dlr_edrs_2016,nasa_smallsat_communications}. Starlink reports ISL bandwidth up to 200~Gbps, while its V3 design specifies 400~Gbps laser terminals~\cite{starlink_progress_2024,starlink_v3_satellites}. The achievable bandwidth of a laser link depends on transmit power, aperture size, pointing accuracy, and link distance. With otherwise identical configurations, longer links increase propagation delay and decrease received optical power, limiting the achievable bandwidth~\cite{esa_ocrsg,jpl_optical_link_characterization}.


\subsection{Related Work}
\label{sec:background-related}

\noindent\textbf{Space-computing infrastructure and systems.}
Google proposes solar-powered TPU satellite clusters with optical interconnects~\cite{google_suncatcher}.
Other studies examine datacenter economics~\cite{orbital_datacenters}, lifecycle emissions~\cite{dirty_bits,llmspace}, and battery aging~\cite{battery_aging}.
Prior work processes space-generated data under compute,
communication, and energy constraints~\cite{oec,space_microdatacenters,architecting_space_microdatacenters,kodan,spacecomp,fool,phoenix,collaborative_inference,orbitserve_vlm}.
Other studies address task and application scheduling in satellite networks ~\cite{leoedge,space_cpn,work_stealing,komet,krios,hyperdrive}.

\noindent\textbf{Orbital LLM systems.}
SpaceMoE~\cite{spacemoe} studies expert placement for a single
MoE replica across multiple satellites and evaluates request-level
token-generation latency, doesn't report
serving-level request-trace or cross-replica scheduling. SkyMemory~\cite{skymemory} compares distributed KV-cache layouts
for a fixed model host.

\noindent\textbf{Terrestrial LLM serving.}
Prior systems optimize resource allocation~\cite{distserve,splitwise}, request migration~\cite{llumnix}, and prefix reuse with load balancing~\cite{srivatsa2025preble}.
SkyWalker~\cite{xia2025skylb} uses locality-aware cross-region routing to balance diurnal demand.
In orbit, scheduling must also account for time-varying ground access and inter-satellite communication paths.

\noindent\textbf{LEO simulation frameworks.}
Existing tools support orbital analysis~\cite{stk}, network simulation~\cite{hypatia,starperf}, software emulation~\cite{celestial}, and onboard-system modeling~\cite{oec}. Differently, OritServe evaluates geo-temporal LLM serving performance and SLO goodput under geo-temporal workloads while accounting for orbital dynamics.
\section{AsterMind Overview}
\label{sec:design}

To explore how orbital compute and communication resources transform into LLM serving performance, AsterMind takes the constellation configuration, geo-temporal workload trace, and runtime policies as inputs. It jointly simulates orbital and runtime dynamics to capture their coupled effects on resource availability and request execution. Finally, AsterMind outputs per-request execution traces and performance metrics, together with the record of session and satellite states.

\begin{figure}[t]
    \centering
    \includegraphics[width=\linewidth]{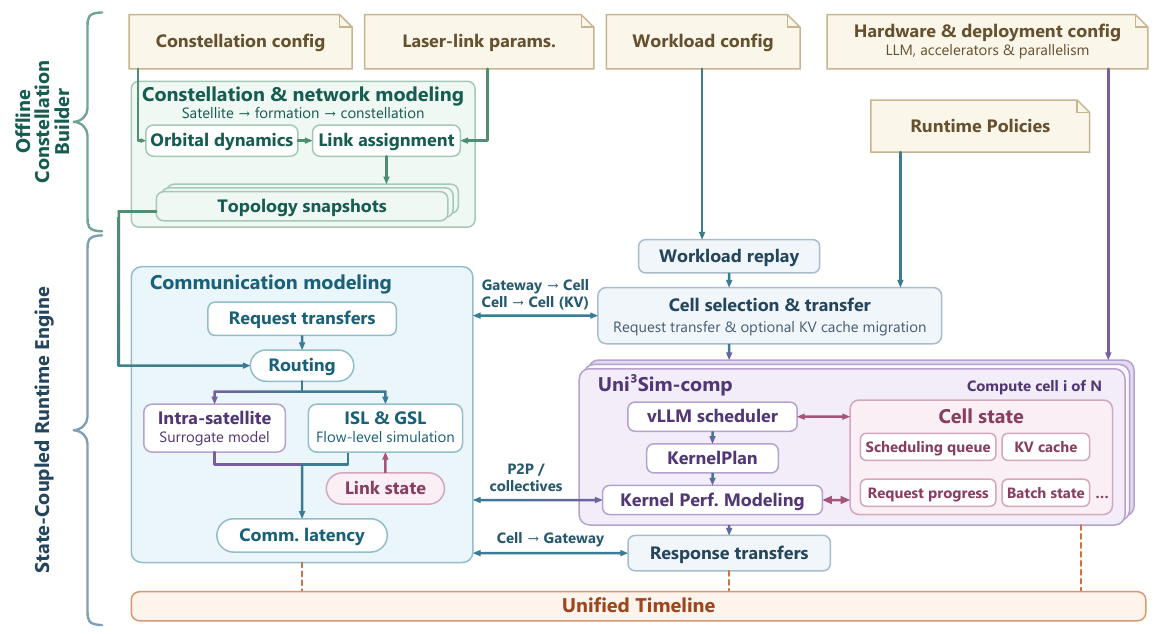}
    \caption{Overview of AsterMind’s architecture for orbital LLM serving. The offline constellation builder prepares the simulation scenario, and the state-coupled runtime engine simulates end-to-end request execution.}
    \label{fig:astermind-overview}
\end{figure}

Figure~\ref{fig:astermind-overview} presents the architecture of AsterMind, whose simulation backend comprises two key components.

\textbf{Offline Constellation Builder.} The builder constructs a hierarchical constellation model from single-satellite, satellite formation to constellation. It parameterizes compute and communication resources within this hierarchy. It propagates satellite orbits, precomputing constellation topology snapshots, candidate communication links, and associated link parameters over the simulation horizon.

\textbf{State-Coupled Runtime Simulation Engine.} The engine replays geo-temporal workload traces on a shared event timeline under configurable deployment, routing, and scheduling policies. It uses the real-time constellation snapshot for routing and communication simulation. It simulates inference for the configured hardware and parallel deployment. The engine simulates communication and inference by coordinating events such as request forwarding, inference batch dispatch and completion, and response delivery. These events update link occupancy, compute queues, and session states, which in turn affect subsequent events. The engine processes these events in chronological order to simulate end-to-end request execution.

\section{Offline Constellation Builder}
\label{sec:offline_constellation_builder}

\subsection{Constellation Modeling}
\label{sec:Constellation_modeling}

\begin{figure}[t!]
    \centering
    \captionsetup{skip=3pt}
    \includegraphics[width=1\linewidth]{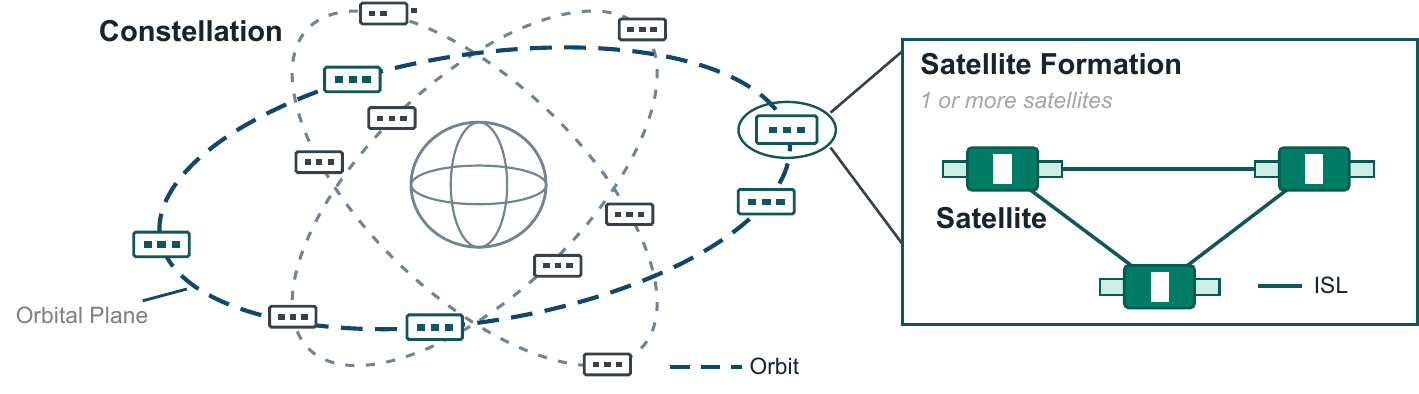}
    \caption{Hierarchical constellation representation.}
    \label{fig:constellation_representation}
\end{figure}

\textbf{Hierarchical constellation representation.} As shown in Figure~\ref{fig:constellation_representation}, AsterMind represents a compute constellation using a three-level hierarchy. At the \emph{Satellite} level, each satellite is instantiated via the parameterized resources. At the \emph{Formation} level, a satellite formation comprises closely spaced satellites. At the \emph{Constellation} level, one or more formations make up a full compute constellation. AsterMind models the constellation topology as a connectivity graph, where nodes represent individual satellites and edges represent established inter-satellite links with distance and bandwidth attributes. 

\paragraph{Orbital Dynamics.}
Satellites in LEO orbits at altitudes of 550--1200~km travel at approximately 7.3--7.6~km/s, completing one orbital period in about 95--110~minutes. This rapid motion continuously changes satellite positions relative to ground gateways, while differences in orbital planes and periods also change the geometry between satellites. The resulting variations in relative distance and viewing angle affect propagation delays, ground-satellite visibility, and inter-satellite line-of-sight (LoS). 

To model these orbital dynamics, AsterMind uses the Simplified General Perturbations 4 (SGP4) propagator~\cite{vallado2006revisiting} to compute satellite positions and velocities at times $t_k=t_0+k\Delta t$, where $\Delta t$ is the sampling interval. The resulting positions are transformed from the True Equator, Mean Equinox (TEME) frame to the Earth-fixed International Terrestrial Reference System (ITRS) for link-geometry calculations. Let $\mathbf r_s(t_k)$ and $\mathbf r_g$ denote the satellite and fixed gateway positions in ITRS, and let $\mathbf n_g$ be the gateway's local upward unit vector. The distance and geometric elevation angle of a ground-satellite link (GSL) are
\begin{equation}
d_{gs}(t_k)=\lVert\mathbf r_s(t_k)-\mathbf r_g\rVert,\qquad \theta_{gs}(t_k)=\arcsin\!\left(\frac{[\mathbf r_s(t_k)-\mathbf r_g]^{\mathsf T}\mathbf n_g}{d_{gs}(t_k)}\right).
\label{eq:astermind-ground-visibility}
\end{equation}
Ground-satellite visibility requires an elevation at or above the minimum threshold $\theta_{g,\min}$ and a distance within the link's operating range. For an inter-satellite link (ISL), the LoS must be unobstructed by the Earth and atmosphere, and the distance must be within the link's operating range. For either link, propagation delay is $\tau=d/c$ for link distance $d$ and light speed $c$. Visibility is evaluated at each sampled time step: $q_e(t_k)=1$ if link $e$ satisfies all applicable physical constraints and $0$ otherwise. These time-varying visibility conditions constrain the active network topology.


\subsection{Network Modeling}
\label{sec:network_modeling}
\begin{figure}[!htbp]
    \centering
    \begingroup
    \setlength{\fboxsep}{5pt}
    \setlength{\fboxrule}{0.45pt}
    \fcolorbox[HTML]{B9C1C7}{FFFFFF}{%
        \begin{minipage}{\dimexpr\linewidth-2\fboxsep-2\fboxrule\relax}
            \begin{minipage}[t]{0.60\linewidth}
                \vspace{0pt}
                \includegraphics[width=\linewidth]{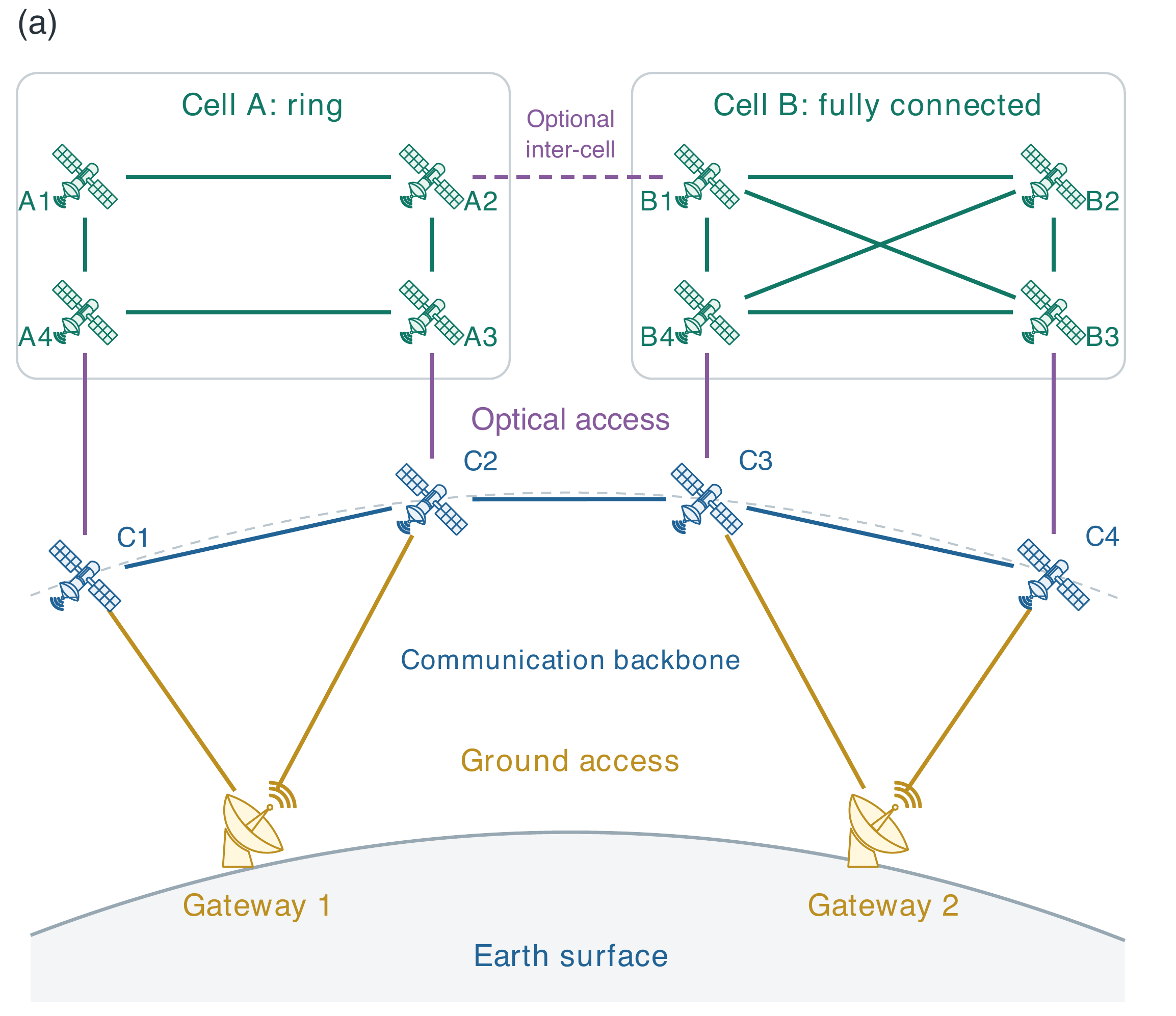}
            \end{minipage}\hfill
            \begin{minipage}[t]{0.38\linewidth}
                \vspace{0pt}
                \includegraphics[width=\linewidth]{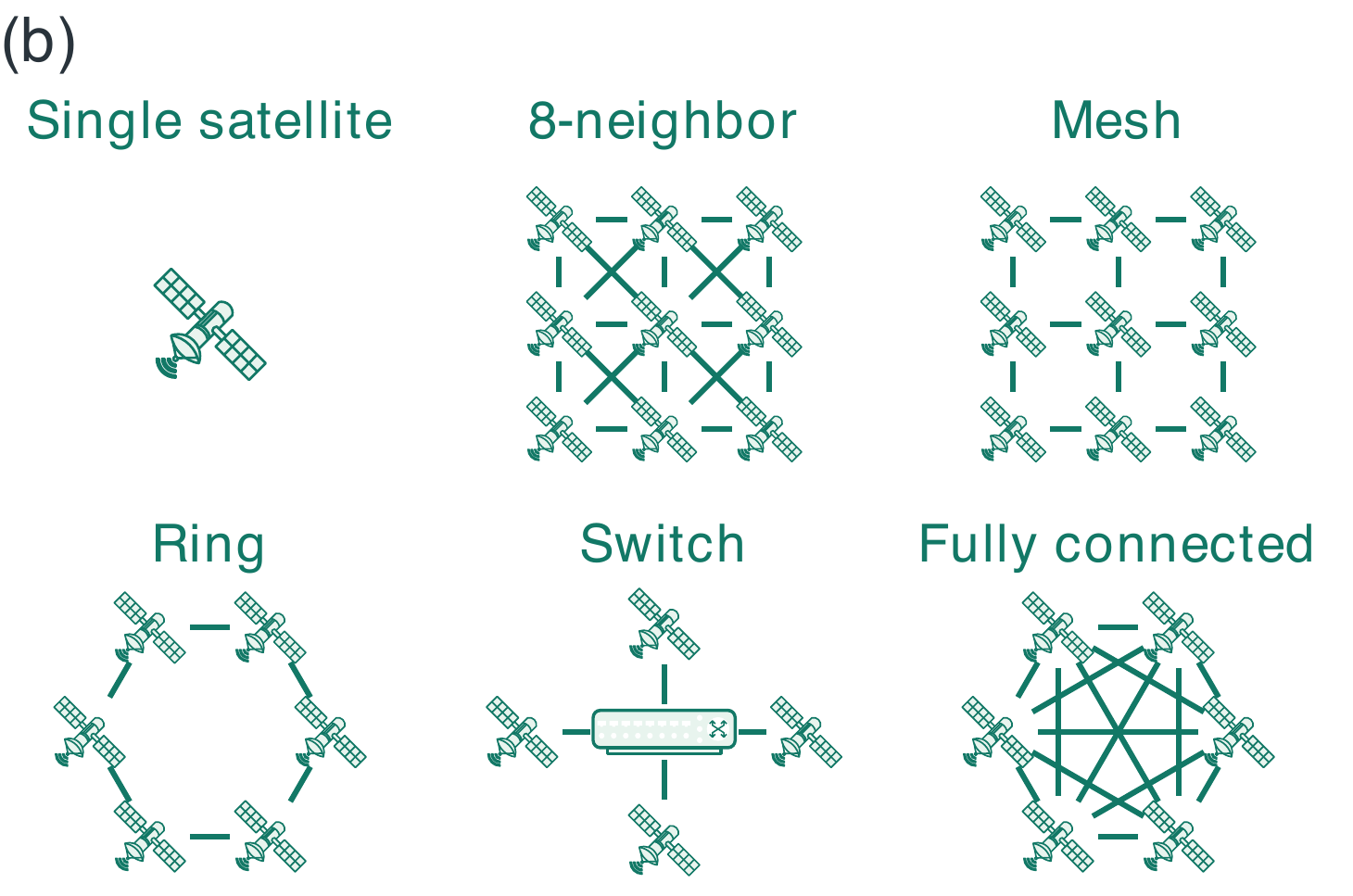}
                \par\vspace{0.025\linewidth}
                {\color[HTML]{D1D6DA}\hrule height 0.4pt}
                \vspace{0.025\linewidth}
                \includegraphics[width=\linewidth]{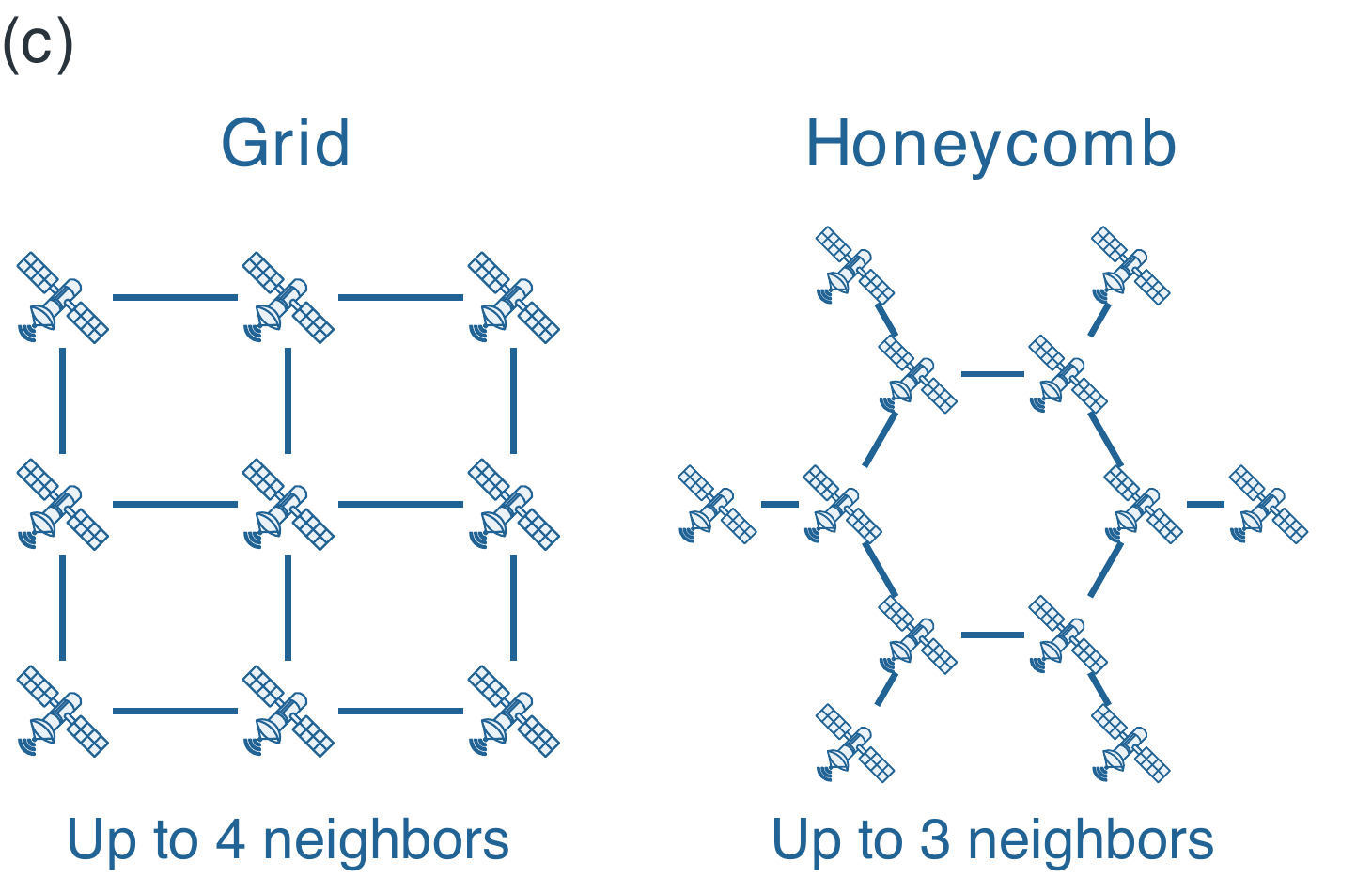}
            \end{minipage}
            \par\vspace{3pt}
            {\color[HTML]{D1D6DA}\hrule height 0.4pt}
            \vspace{3pt}
            {\footnotesize
                \textcolor[HTML]{226396}{Communication satellite}\hfill
                \textcolor[HTML]{127866}{Compute satellite}\hfill
                \textcolor[HTML]{127866}{\rule[0.4ex]{1.5em}{0.6pt}}\enspace Intra-cell\hfill
                \textcolor[HTML]{865A9C}{\rule[0.4ex]{0.4em}{0.6pt}\kern0.2em\rule[0.4ex]{0.4em}{0.6pt}\kern0.2em\rule[0.4ex]{0.4em}{0.6pt}}\enspace Inter-cell\par
            }
        \end{minipage}%
    }
    \endgroup
    \caption{AsterMind's network model: (a) Network architecture connecting ground gateways, communication satellites and compute satellites; (b) Compute cell topologies; (c) Communication topologies, including the 4-ISL Grid and 3-ISL Honeycomb.}
    \label{fig:astermind-network-topology}
\end{figure}

The network comprises ground gateways, communication satellites, and formations of compute satellites, referred to as compute cells (Figure~\ref{fig:astermind-network-topology}(a)). The communication constellation supports a 3-ISL Honeycomb~\cite{chen2025threeisl} or 4-ISL Grid~\cite{bhattacherjee2019topology} topology (Figure~\ref{fig:astermind-network-topology}(c)). These ISLs form the communication backbone that relays data between ground users and compute nodes. Within a multi-satellite formation, the intra-cell topology can be configured as 8-neighbor, mesh, ring, switch-based, or fully connected (Figure~\ref{fig:astermind-network-topology}(b)).

For ground access, gateways establish GSLs with satellites selected according to distance, elevation, available capacity, or remaining contact time. Communication-to-compute access uses optical links between satellite pairs, selected by the same criteria except elevation. At each time step, selection is restricted to physically feasible pairs and respects terminal limits at both nodes. Two connection policies are supported: Stability-First retains valid links and uses the configured ranking to fill free terminals, whereas Score-First re-evaluates dynamic link assignments according to the current ranking. Newly selected links become active only after their configured acquisition time, accounting for link-setup overhead~\cite{bhattacherjee2019topology}. Inter-cell transfers can traverse the communication backbone or use direct links between compute satellites in different cells.

Let $\mathcal N$ be the set of all gateways and satellites, and let $\mathcal E_{\mathrm{cand}}$ contain the candidate links allowed by the configured topologies and access rules. At sampled time $t_k$, the active network is
\begin{equation}
G(t_k)=(\mathcal N,\mathcal E(t_k)),\qquad \mathcal E(t_k)=\{e\in\mathcal E_{\mathrm{cand}}\mid q_e(t_k)=1,\ a_e(t_k)=1\},
\label{eq:astermind-active-network}
\end{equation}
where $\mathcal E(t_k)$ is the active link set. For each candidate link $e$, $q_e(t_k)=1$ if all applicable physical constraints, including elevation, LoS, and operating range, are satisfied; $a_e(t_k)=1$ if the required terminals are allocated and acquisition is complete. Each indicator is $0$ otherwise.

\subsection{Fault Injection}
\label{sec:builder-fault-injection}

An energetic particle can trip the power controller supplying a satellite's communications equipment, interrupting both transmission and reception~\cite{image2006failure}. Reaction-wheel vibrations can induce pointing errors in an optical terminal, steering its beam off target and reducing the received optical power below the detection threshold~\cite{thiele2023vibration,hayes1997fading}. AsterMind models these outages as node faults and link faults, respectively.

AsterMind supports two fault models. The \emph{scheduled fault model} randomly selects a fraction of satellites or links and disables all selected targets during the same user-defined time window. The \emph{random fault model} generates a separate sequence of outages for each target by sampling its operating times and assigning each outage a fixed or sampled duration.

At time $t$ in the scenario, let $G(t)$ denote the fault-free graph with node set $\mathcal N$ and the links active at that time in the precomputed topology. At each sampling instant $t_k$, Equation~\eqref{eq:astermind-active-network} defines $G(t_k)$. Let $\mathcal{N}_{\mathrm{f}}(t)$ denote the satellites whose fault intervals contain $t$, and let $\mathcal{E}_{\mathrm{f}}(t)$ denote the candidate links with an ongoing outage of their own. We use $-$ to denote graph deletion. Deleting a node removes its incident links as well, while deleting a link preserves its endpoints. The available graph $G_{\mathrm{avail}}(t)$ is
\begin{equation}
G_{\mathrm{avail}}(t)
= \bigl(G(t) - \mathcal{N}_{\mathrm{f}}(t)\bigr)
  - \mathcal{E}_{\mathrm{f}}(t).
\label{eq:builder-fault-availability}
\end{equation}

\subsection{Laser-Link Parameterization}
\label{sec:builder-laser-links}

The constellation builder maps constellation topology and
optical-terminal parameters to per-link capacity,
propagation delays, and availability. Link distance $d_{ij}$ between satellite $i$ and $j$ is obtained from user-defined constellation topology, including user-defined formation spacing or satellite
positions obtained from orbital propagation.
Together with transmit power, aperture diameters, and
transmission parameters, AsterMind evaluates the optical link budget. Alternatively, users can also specify link capacity directly.

For each wavelength channel, the received optical power
is estimated as:
\begin{equation}
\begin{aligned}
P_r
&=
P_{t,\mathrm{ch}} G_t G_r
\left(\frac{\lambda}{4\pi d_{ij}}\right)^2
\eta_{\mathrm{opt}}\eta_{\mathrm{point}}
10^{-L_{\mathrm{add}}/10}, &
G_{t,r}
=
\left(\frac{\pi D_{t,r}}{\lambda}\right)^2,
\end{aligned} 
\label{eq:builder-optical-budget}
\end{equation}
where $P_{t,\mathrm{ch}}$ is the transmit power per-channel,
$\lambda$ is the optical wavelength, and $D_t$ and $D_r$
are the aperture diameters of the transmitter and receiver.
The factors $\eta_{\mathrm{opt}}$ and
$\eta_{\mathrm{point}}$ account for optical efficiency
and pointing attenuation.
The additional loss is $L_{\mathrm{add}}=L_{\mathrm{margin}}+L_{\mathrm{WDM}}$,
where $L_{\mathrm{margin}}$ and $L_{\mathrm{WDM}}$ are the link margin and wavelength-division multiplexing (WDM) loss in decibels.
Under the default equal-power allocation, $P_{t,\mathrm{ch}}=P_t/N_{\lambda}$, where $P_t$ is the total transmit power and $N_{\lambda}$ is the number of wavelength channels.
Pointing attenuation is $\eta_{\mathrm{point}} =\exp[-2(\theta_{\mathrm{err}}/\theta_b)^2]$, where $\theta_b=\lambda/(\pi w_0)$, $w_0$ is the beam-waist radius, and $\theta_{\mathrm{err}}$ is the pointing error.


For each evaluated symbol rate $R_s$, the per-polarization
signal-to-noise ratio (SNR) is
\begin{equation}
    \gamma(R_s)
    =
    \frac{P_r}
    {N_{\mathrm{pol}} F_n h \nu R_s},
    \quad
    F_n = 10^{\mathrm{NF}/10},
    \quad
    \nu = \frac{c}{\lambda},
    \label{eq:laser-snr}
\end{equation}
where $F_n$ is the linear preamplifier noise factor,
$h$ is Planck's constant, and $\nu$ is the optical frequency.
This model assumes equal power allocation across
polarizations and signal--amplified spontaneous emission
(ASE) beat noise dominance under the high-gain approximation.

The implemented bit-error-rate (BER) approximation is
\begin{equation}
    \widehat{\mathrm{BER}}(M,R_s)
    =
    \frac{2(\sqrt{M}-1)}{\sqrt{M}\log_2 M}
    Q\!\left(
        \sqrt{\frac{3\gamma(R_s)}{M-1}}
    \right),
    \label{eq:laser-ber}
\end{equation}
where $M$ is the modulation order and $Q(\cdot)$ is the
standard Gaussian tail function.
At each rate, the model selects the first format in
$\mathcal{M}$ satisfying
$\widehat{\mathrm{BER}}\leq\mathrm{BER}_{\mathrm{th}}$.
Fixed-rate mode uses the configured $R_s$.
Adaptive mode evaluates $\mathcal{R}_s$ and selects the
feasible rate--modulation pair with the highest capacity
after applying the terminal limit.

For a feasible symbol-rate and modulation pair, the net
physical capacity is
\begin{equation}
C_{\mathrm{phy}}
=
N_{\lambda}N_{\mathrm{pol}}R_s\log_2(M)
(1-f_{\mathrm{FEC}}),
\label{eq:laser-physical-capacity}
\end{equation}
where $N_{\lambda}$ and $N_{\mathrm{pol}}$ are the numbers
of wavelength and polarization channels, respectively,
$R_s$ is the symbol rate, $M$ is the modulation order, and
$f_{\mathrm{FEC}}$ is the fractional capacity deduction
for forward error correction (FEC). The default configuration uses a fixed symbol rate of
60~GBaud. Adaptive operation evaluates the configured rate candidates and selects the largest capacity after applying the terminal limit. The link budget is calculated as:
\begin{equation}
C^{\mathrm{budget}}
=
\min\!\left(
C_{\mathrm{phy}},
C_{\mathrm{term}},
C_{\mathrm{profile}}
\right),
\label{eq:laser-budget-capacity}
\end{equation}
where $C_{\mathrm{term}}$ is the optical-terminal capacity
limit and $C_{\mathrm{profile}}$ is the configured
link-profile limit.
A disabled terminal limit is treated as unbounded. If no evaluated transmission format satisfies the BER threshold, the link budget is set to zero. Table~\ref{tab:builder-laser-defaults} summarizes the reference optical-terminal parameters and default waveform settings, which are configurable for individual link profiles.

\begin{table}[t]
  \centering
  \setlength{\abovecaptionskip}{0pt}
  \setlength{\belowcaptionskip}{8pt}
  \caption{Reference configuration for inter-satellite laser-link parameterization.}
  \label{tab:builder-laser-defaults}
  \small
  \setlength{\tabcolsep}{5pt}
  \renewcommand{\arraystretch}{1.12}

  \begin{tabular}{@{}
    >{\raggedright\arraybackslash}p{6.5cm}
    >{\centering\arraybackslash}p{1.6cm}
    >{\raggedright\arraybackslash}p{4cm}
  @{}}
    \toprule
    \textbf{Parameter}
      & \textbf{Symbol}
      & \textbf{Reference value} \\
    \midrule

    \multicolumn{3}{@{}l}{\itshape Optical terminals} \\
    \addlinespace[2pt]
    Total optical transmit power
      & $P_t$ & 5~W \\
    Transmitter aperture diameter
      & $D_t$ & 0.10~m \\
    Receiver aperture diameter
      & $D_r$ & 0.10~m \\
    Nominal optical wavelength
      & $\lambda$ & 1550~nm \\
    Beam-waist radius
      & $w_0$ & 0.05~m \\
    Optical efficiency
      & $\eta_{\mathrm{opt}}$ & 0.7 \\
    Pointing error
      & $\theta_{\mathrm{err}}$ & 0~rad \\
    Link margin
      & $L_{\mathrm{margin}}$ & 7~dB \\
    WDM loss
      & $L_{\mathrm{WDM}}$ & 6~dB \\

    \addlinespace[6pt]
    \multicolumn{3}{@{}l}{\itshape Transmission and receiver settings} \\
    \addlinespace[2pt]
    Wavelength channels
      & $N_{\lambda}$ & 2 \\
    Polarization channels per wavelength
      & $N_{\mathrm{pol}}$ & 2 \\
    Fixed symbol rate
      & $R_s$ & 60~GBaud \\
    Candidate symbol rates (adaptive mode)
      & $\mathcal{R}_s$
      & $\{20,30,40,50,60\}$~GBaud \\
    Modulation-order priority
      & $\mathcal{M}$
      & $16\rightarrow8\rightarrow4$ \\
    Optical preamplifier noise figure
      & $\mathrm{NF}$ & 7.5~dB \\
    BER threshold
      & $\mathrm{BER}_{\mathrm{th}}$ & $10^{-3}$ \\
    FEC capacity deduction
      & $f_{\mathrm{FEC}}$ & 15\% \\


    \bottomrule
  \end{tabular}
\end{table}


\subsection{Hardware and Deployment Configuration}
\label{sec:design-compute}
At initialization, users specify the LLM, accelerator type, and parallelism strategy for each compute cell, along with the deployment mappings from the model to accelerators and from accelerators to satellites. AsterMind loads the hardware profile associated with the selected accelerator type. Table~\ref{tab:parallelism-coverage} reports end-to-end simulation coverage for different LLMs and parallelism strategies on four accelerator types.


\begin{table*}[t]
  \centering
  \caption{End-to-end simulation coverage across models, parallelism strategies, and four accelerator types (NVIDIA H200 / NVIDIA A800 / MetaX C500 / Ascend 910B)}
  \label{tab:parallelism-coverage}
  \small
  \setlength{\tabcolsep}{4pt}
  \renewcommand{\arraystretch}{1.12}

  \providecommand{\OSCoverPass}{\ensuremath{\checkmark}}
  \providecommand{\OSCoverCross}{\ensuremath{\times}}
  \newcommand{\OSCoverHBM}{\textnormal{X}}
  \newcommand{\OSCover}[4]{%
    #1\kern.04em/\kern.04em#2\kern.04em/\kern.04em#3\kern.04em/\kern.04em#4}
  \newcommand{\OSCoverAll}{%
    \OSCover{\OSCoverPass}{\OSCoverPass}{\OSCoverPass}{\OSCoverPass}}
  \newcommand{\OSCoverNA}{%
    \OSCover{\mbox{--}}{\mbox{--}}{\mbox{--}}{\mbox{--}}}

  \resizebox{\linewidth}{!}{%
    \begin{tabular}{@{}lccccccc@{}}
      \toprule
      \multirow{2}{*}[-3pt]{\textbf{Model}}
        & \multicolumn{5}{c}{$\mathrm{DP}=1$}
        & \multicolumn{2}{c}{$\mathrm{DP}>1$} \\
      \cmidrule(lr){2-6}\cmidrule(lr){7-8}
        & \textbf{TP}
        & \textbf{PP}
        & \textbf{TP\OSCoverCross{}PP}
        & \textbf{TP\OSCoverCross{}EP}
        & \textbf{TP\OSCoverCross{}PP\OSCoverCross{}EP}
        & \textbf{WideEP}
        & \textbf{WideEP\OSCoverCross{}PP} \\
      \midrule
      DeepSeek-V3
        & \OSCoverAll & \OSCoverAll & \OSCoverAll & \OSCoverAll
        & \OSCoverAll & \OSCoverAll & \OSCoverAll \\
      DeepSeek-V3-L32
        & \OSCoverAll & \OSCoverAll & \OSCoverAll & \OSCoverAll
        & \OSCoverAll & \OSCoverAll & \OSCoverAll \\
      DeepSeek-V4-Flash
        & \OSCoverAll & \OSCoverAll & \OSCoverAll & \OSCoverAll
        & \OSCoverAll & \OSCoverAll & \OSCoverAll \\
      DeepSeek-V4-Pro
        & \OSCover{\OSCoverPass}{\OSCoverPass}{\OSCoverHBM}{\OSCoverHBM}
        & \OSCoverAll & \OSCoverAll & \OSCoverAll
        & \OSCoverAll & \OSCoverAll & \OSCoverAll \\
      GLM-5.2
        & \OSCoverAll
        & \OSCover{\OSCoverPass}{\OSCoverHBM}{\OSCoverHBM}{\OSCoverHBM}
        & \OSCoverAll & \OSCoverAll & \OSCoverAll
        & \OSCoverAll & \OSCoverAll \\
      GLM-5.3
        & \OSCoverAll & \OSCoverAll & \OSCoverAll & \OSCoverAll
        & \OSCoverAll & \OSCoverAll & \OSCoverAll \\
      Kimi K2.6
        & \OSCoverAll & \OSCoverAll & \OSCoverAll & \OSCoverAll
        & \OSCoverAll & \OSCoverAll & \OSCoverAll \\
      Kimi K2.7-Code
        & \OSCoverAll & \OSCoverAll & \OSCoverAll & \OSCoverAll
        & \OSCoverAll & \OSCoverAll & \OSCoverAll \\
      Kimi K3
        & \OSCover{\OSCoverHBM}{\OSCoverHBM}{\OSCoverHBM}{\OSCoverHBM}
        & \OSCoverAll & \OSCoverAll
        & \OSCover{\OSCoverPass}{\OSCoverPass}{\OSCoverHBM}{\OSCoverHBM}
        & \OSCoverAll & \OSCoverAll & \OSCoverAll \\
      Qwen3-30B-A3B
        & \OSCoverAll & \OSCoverAll & \OSCoverAll & \OSCoverAll
        & \OSCoverAll & \OSCoverAll & \OSCoverAll \\
      \midrule
      Llama-3.1-8B-Instruct
        & \OSCoverAll & \OSCoverAll & \OSCoverAll
        & \OSCoverNA & \OSCoverNA & \OSCoverNA & \OSCoverNA \\
      Qwen3-32B
        & \OSCoverAll & \OSCoverAll & \OSCoverAll
        & \OSCoverNA & \OSCoverNA & \OSCoverNA & \OSCoverNA \\
      Qwen3-32B (FP16)
        & \OSCoverAll & \OSCoverAll & \OSCoverAll
        & \OSCoverNA & \OSCoverNA & \OSCoverNA & \OSCoverNA \\
      Qwen3-4B
        & \OSCoverAll & \OSCoverAll & \OSCoverAll
        & \OSCoverNA & \OSCoverNA & \OSCoverNA & \OSCoverNA \\
      Qwen3-8B
        & \OSCoverAll & \OSCoverAll & \OSCoverAll
        & \OSCoverNA & \OSCoverNA & \OSCoverNA & \OSCoverNA \\
      \bottomrule
    \end{tabular}%
  }

  \par\smallskip
  \begin{minipage}{\linewidth}
    \small
      \OSCoverPass{} indicates that at least one tested configuration
      completed an end-to-end request simulation.\\
      \OSCoverHBM{} indicates that the tested placements exceeded the per-device HBM budget.\\
      \mbox{--} indicates that the strategy is inapplicable.%

  \end{minipage}
\end{table*}


\section{State-Coupled Runtime Engine}
\label{sec:runtime_engine}

AsterMind replays geo-temporal workloads and advances request execution on a shared event timeline, coupling communication, compute, and session states. At each event, the engine evaluates execution dependencies and resource availability using the current system state to determine which operations can proceed. The resulting state updates affect subsequent events, capturing how ongoing transfers and LLM inference affect resource contention and the progress of other requests.

For each arriving request, configured runtime policies determine request routing and scheduling based on the active constellation topology. The compute model captures prefill and decode execution under the specified hardware and parallelism configuration, accounting for batching and KV-cache reuse. The communication model determines when the required data becomes available for inference and response transfers over the evolving network, accounting for link availability, propagation delays and bandwidth contention.



\subsection{Workload Replay}
\label{sec:workload_replay}
\begin{figure}[t]
    \centering
    \captionsetup{skip=3pt}
    \includegraphics[width=1\linewidth]{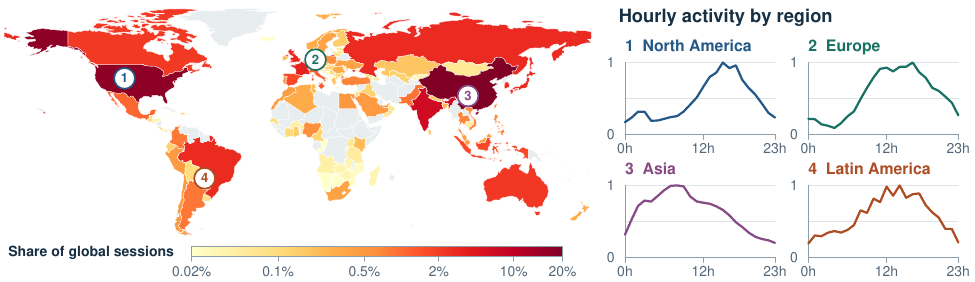}
    \caption{Global LLM workload characteristics.}
    \label{fig:workload}
\end{figure}

AsterMind loads standardized workload bundles, each containing request and session identifiers, turn indices, ingress gateways, arrival times, and input/output lengths. For each request, the replayer generates a request arrival event using its specified time and ground gateway on the shared event timeline. This preserves regional differences in request volume and diurnal request patterns encoded in the bundles, as illustrated in Figure~\ref{fig:workload}. The workload replayer also preserves session associations and turn ordering, which identify the context available for KV-cache reuse across turns within the same session. Users can configure workload intensity to explore different demand levels. Fixing the replay configuration and random seed yields a reproducible request stream.

\subsection{Runtime Policies}
\label{sec:runtime_policies}
AsterMind exposes runtime policies for request scheduling, routing and session management. At request arrival, AsterMind selects a compute cell according to the scheduling policy. The \textit{least-loaded policy} prioritizes cells with the fewest outstanding requests followed by the shortest ingress distance. The \textit{nearest policy} reverses this order by prioritizing distance first. The \textit{estimated-arrival policy} prioritizes outstanding request count, but breaks ties using estimated input-transfer time.

Routing operates on the active network graph and employs \textit{min-hops}, \textit{min-propagation}, or \textit{user-defined routing algorithms}. Data transfers in progress adapt to changes in link capacity. If a transfer's path becomes unavailable, AsterMind triggers an end-to-end retransmission between the original source and destination. Collective communication operations are mapped onto physical routes, with link capacities shared through max-min fair allocation.

For each request, AsterMind checks for a reusable KV-cache prefix from earlier turns of the same session, including prefixes stored on other compute cells. If the selected cell lacks the required cache but a reusable prefix is available remotely, the runtime engine supports \textit{KV-cache migration}. When migration is enabled, missing reusable KV-cache is transferred only if the predicted migration cost is lower than that of context recomputation; otherwise, the missing context is recomputed locally. Once the migration completes, the destination cell updates its session-level cache state, making the migrated prefix available for reuse. 





\subsection{Communication Modeling}
\label{sec:communication_modeling}
\begin{figure}[t!]
    \centering
    \includegraphics[width=\linewidth]{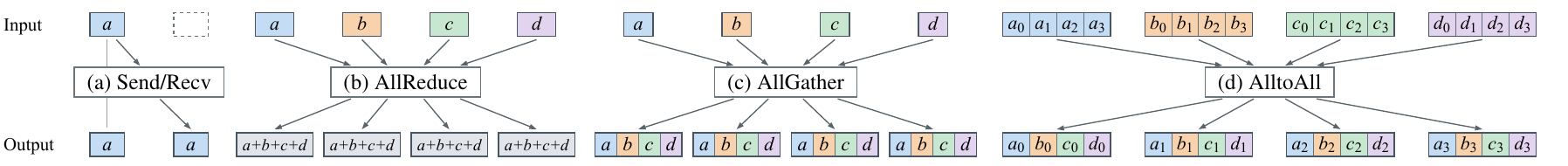}
    \caption{Data layout across ranks before and after Send/Recv, AllReduce, AllGather, and AlltoAll.}
    \label{fig:communication-operators}
\end{figure}

Evaluating constellation-scale LLM serving requires tracking concurrent request transfers, KV-cache migrations, and collective communications in LLM inference over time-varying network topology. Packet-level simulator like NS-3~\cite{riley2010ns} processes individual packet events, increasing simulation cost as traffic volume grows. To address this issue, AsterMind combines protocol-aware latency estimation with flow-level resource sharing for communication modeling. Figure~\ref{fig:communication-operators} illustrates four communication
operators. AsterMind maps the transfers required by each operator to physical paths and executes them according to their dependencies. For the inter-rank communication tasks, the physical placement of the participating ranks determines whether a transfer uses the intra-satellite accelerator interconnects or traverses inter-satellite links. The communication model selects the appropriate timing backend based on the physical path and the configured policy.

\textbf{Protocol-aware surrogate model.}
To capture protocol effects without simulating individual packet events, AsterMind adopts the analytical
surrogate model from ProtoBridge~\cite{protobridge}:

\begin{equation}
\begin{aligned}
\widehat{L} ={}& S_p + k_a T_{\mathrm{fixed}} + k_a T_{\mathrm{serial}}\Phi_{\mathrm{credit}}\Phi_{\mathrm{fec}}\Phi_{\mathrm{share}} + T_{\mathrm{retx}} + T_{\mathrm{reload}} + T_{\mathrm{fec,lat}}.
\end{aligned}
\label{eq:protocol-surrogate}
\end{equation}

Here, $S_p$ is the protocol startup cost, and $k_a$ is the number of
sequential transfer steps determined by the communication algorithm.
$T_{\mathrm{fixed}}$ and $T_{\mathrm{serial}}$ denote the fixed delay and baseline serialization time per step, respectively. The factors $\Phi_{\mathrm{credit}}$, $\Phi_{\mathrm{fec}}$, and
$\Phi_{\mathrm{share}}$ account for credit limitations, FEC coding
expansion, and shared-resource contention. The remaining terms represent retransmission, source-data reload, and FEC encoding and decoding latency.

\paragraph{Flow-level inter-satellite simulation.} Across satellites, concurrent transfers compete for link capacity over an evolving network. Capturing these interactions requires tracking transfer progress as connectivity and resource availability change. AsterMind therefore uses a flow-level simulator to evaluate request forwarding, KV-cache migration, and inter-satellite communication during distributed LLM inference. Concurrent flows share link capacities through max-min fair allocation~\cite{nguyen2009efficient}, with their service rates updated as link capacities and competing traffic change. We adapt the protocol-aware surrogate model to the laser-link profiles, including laser-link capacity, framing and error-control parameters. As the laser-link parameterization introduced in Section~\ref{sec:builder-laser-links} already incorporates the configured FEC overhead, $\Phi_{\mathrm{fec}}$ is set to $1$ and $T_{\mathrm{fec,lat}}$ remains a separate term. When flow-level simulation is enabled, we set $\Phi_{\mathrm{share}}$ to 1 because the flow simulator already accounts for bandwidth sharing. Note that the above communication profiles are configurable and alternative models can be integrated into AsterMind. 


\subsection{Inference Simulation}
\label{sec:fast-batch-simulation}

AsterMind supports two modes for batch execution, \emph{Detailed} mode and \emph{Fast} mode. Both use the same vLLM~\cite{vllm} scheduler, which supports continuous batching, chunked prefill, and KV-cache reuse. Once the input transfer and any required KV-cache transfers are ready, the request enters the scheduling queue of its assigned compute cell. At each scheduling step, vLLM groups eligible prefill and decode work into a batch, which denotes a set of requests processed together in one model forward pass. Both modes record one \emph{batch completion event} on the shared timeline when all computation and inter-rank communication for the batch have finished. AsterMind then updates request progress and KV-cache state for subsequent batches.


\paragraph{Detailed Mode.}
For each scheduled batch, AsterMind uses Uni\textsuperscript{3}Sim-comp\footnote{The Uni\textsuperscript{3}Sim repository is available at {\urlstyle{same}\url{https://github.com/thu-nics/uni3sim}}.}~\cite{uni3sim} to construct a \textit{WorkloadGraph} that specifies per-rank operations, tensor shapes, inter-rank communication, and execution dependencies. It then lowers the graph into a \textit{KernelPlan} comprising compute kernel invocations and communication tasks. Uni\textsuperscript{3}Sim-comp estimates the duration of each compute kernel.

When a communication task becomes ready, AsterMind maps the participating ranks to the deployed devices and submits the operation type, participating devices, and message size to the communication model described in Section~\ref{sec:communication_modeling}. Communication completion enables dependent tasks to proceed. Computation and communication progress on the shared event timeline as dependencies and resources permit. One batch completes when all of its compute and communication tasks have finished.

\paragraph[Fast Mode.]{Fast Mode.}
At deployment, AsterMind prepares a forward execution plan using Uni\textsuperscript{3}Sim's \textit{WorkloadGraph} as a reference. The plan specifies the compute steps for each layer type and where inter-rank communication occurs. For each batch scheduled by vLLM, AsterMind uses the scheduled tokens and request contexts to estimate the durations of different layer types, including projections, attention, and feed-forward networks (FFNs). 

\begin{figure}[H]
    \centering
    \captionsetup{skip=3pt}
    \includegraphics[width=0.85\linewidth]{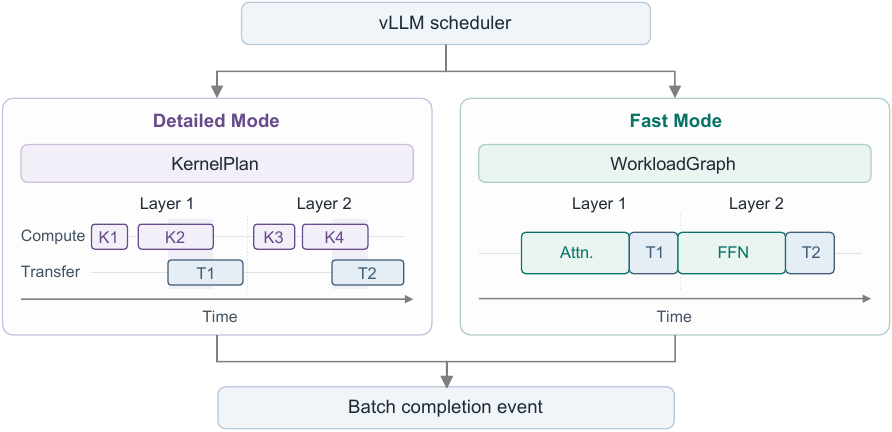}
    \caption{AsterMind’s two modes for inference simulation.}
    \label{fig:batch-execution}
\end{figure}

For each of these compute steps, AsterMind uses the Roofline model to estimate latency as the maximum of the compute time and the HBM access time.
\begin{equation}
    T = \max\left(\frac{W}{P_{\mathrm{eff}}},\,\frac{Q}{B_{\mathrm{eff}}}\right).
    \label{eq:fast-batch-roofline}
\end{equation}
Here, $W$ and $Q$ are the floating-point operation count and HBM traffic in bytes. $P_{\mathrm{eff}}$ and $B_{\mathrm{eff}}$ are the effective compute throughput and effective HBM bandwidth, respectively. Both modes use the protocol-aware surrogate described in Section~\ref{sec:communication_modeling} for intra- and inter-satellite communication. For inter-satellite transfers, \emph{Detailed} mode models bandwidth contention through flow-level simulation, while the \emph{Fast} mode approximates bandwidth contention with a user-configured sharing factor $\Phi_{\mathrm{share}}$ (Eq.~\eqref{eq:protocol-surrogate}). As shown in Figure~\ref{fig:batch-execution}, the \emph{Fast} mode treats computation and communication as non-overlapping, sums the estimated durations of all steps, and avoids building a \textit{KernelPlan} for each batch.




\section{Validation}
\textbf{Orbital model validation.}
We validate AsterMind's orbital model against the Ansys Systems Tool Kit (STK)~\cite{stk} using identical satellite definitions, epochs, ground-station locations, elevation masks, and sampling intervals. For each of four representative satellites, we compare the satellite position and GSL access windows simulated by STK and AsterMind. Table~\ref{tab:orbital_model_validation} reports the average position error, defined as the average Euclidean distance between corresponding positions across all sampled timestamps, and the intersection over union (IoU) of the GSL access windows. Across the evaluated satellites, AsterMind achieves average position errors below 0.00049~m and GSL window IoUs of at least 99.17\%. The results further validate the high fidelity of AsterMind's orbital modeling.

\begin{table}[H]
\centering
\captionsetup{skip=3pt}
\renewcommand{\arraystretch}{0.85}
\caption{Validation of AsterMind's orbital positions and GSL access windows against STK.}
\label{tab:orbital_model_validation}
\small
\setlength{\tabcolsep}{5pt}
\begin{tabular}{ccc}
\toprule
Satellite
& \shortstack{Avg. Position Error (m) $\downarrow$}
& \shortstack{GSL Window IoU (\%) $\uparrow$} \\
\midrule
10000 & 0.000479 & 99.77 \\
10021 & 0.000489 & 99.33 \\
10063 & 0.000475 & 99.76 \\
10084 & 0.000475 & 99.17 \\
\bottomrule
\end{tabular}
\end{table}

\textbf{Communication model validation.} 
We compare AsterMind with ns-3 for four communication operators across five message sizes from 4 KiB to 64 MiB, with matched workloads, topology and link parameters. The intra-satellite tests use eight GPUs within a single satellite, whereas the inter-satellite
tests use one GPU per satellite in a eight-satellite ring with 1 km inter-satellite links. Table~\ref{tab:comm-validation-intra} shows that AsterMind estimates intra-satellite communication latency with average relative errors (AREs) ranging from 5.01\% to 13.91\%. 

Then we evaluate the simulation errors when GPUs are distributed across satellites. The evaluation is conducted on an eight-satellite ring with one GPU per satellite and 1~km links.
As shown in Table~\ref{tab:comm-validation-inter}, across the five message sizes, the ARE is 10.78\% for Send/Recv, 12.25\% for AllReduce, and 10.78\% for AllGather. Pairwise AlltoAll spans up to four hops and has an ARE of 47.71\%. In this configuration, the largest simulation discrepancy occurs for AlltoAll at 64~MiB per GPU, where it predicts 2.88~ms compared with 1.31~ms in ns-3.

\begin{table}[H]
    \centering
    \caption{Simulation of intra-satellite communication latency.
    Latencies are reported in $\mu$s.}
    \label{tab:comm-validation-intra}

    \begingroup
    \small
    \setlength{\tabcolsep}{2pt}
    \renewcommand{\arraystretch}{1}

    \resizebox{\linewidth}{!}{%
        \begin{tabular}{@{}l*{10}{c}r@{}}
            \toprule
            \multirow{2}{*}{Operator}
            & \multicolumn{2}{c}{4 KiB}
            & \multicolumn{2}{c}{64 KiB}
            & \multicolumn{2}{c}{1 MiB}
            & \multicolumn{2}{c}{16 MiB}
            & \multicolumn{2}{c}{64 MiB}
            & \multirow{2}{*}{\mbox{ARE(\%)}} \\
            \cmidrule(lr){2-3}
            \cmidrule(lr){4-5}
            \cmidrule(lr){6-7}
            \cmidrule(lr){8-9}
            \cmidrule(lr){10-11}
            & \commcolhead{AsterMind} & \commcolhead{ns-3}
            & \commcolhead{AsterMind} & \commcolhead{ns-3}
            & \commcolhead{AsterMind} & \commcolhead{ns-3}
            & \commcolhead{AsterMind} & \commcolhead{ns-3}
            & \commcolhead{AsterMind} & \commcolhead{ns-3}
            & \\
            \midrule
            Send/Recv & 15.04 & 15.30 & 15.39 & 15.92 & 20.95 & 21.44 & 109.81 & 99.84 & 394.17 & 351.75 & 5.87 \\
            AllReduce & 15.32 & 18.22 & 15.93 & 19.31 & 25.65 & 32.78 & 181.16 & 174.21 & 678.79 & 615.04 & 13.91 \\
            AllGather & 15.30 & 17.11 & 17.73 & 21.44 & 56.61 & 60.08 & 678.65 & 608.91 & 2669.16 & 2390.63 & 11.35 \\
            AlltoAll & 46.16 & 47.61 & 46.46 & 48.16 & 51.32 & 54.89 & 129.08 & 125.60 & 377.89 & 346.02 & 5.01 \\
            \bottomrule
        \end{tabular}%
    }
    \endgroup

    \par\smallskip
    \begin{minipage}{\linewidth}
        \footnotesize
        ARE: average relative error versus ns-3 across
        the five message sizes.
    \end{minipage}
\end{table}

\begin{table}[H]
    \centering
    \caption{Inter-satellite communication on an eight-satellite ring
    with one GPU per satellite.}
    \label{tab:comm-validation-inter}

    \begingroup
    \small
    \setlength{\tabcolsep}{2pt}
    \renewcommand{\arraystretch}{1}

    \resizebox{\linewidth}{!}{%
        \begin{tabular}{@{}l*{10}{c}r@{}}
            \toprule
            \multirow{2}{*}{Operator}
            & \multicolumn{2}{c}{4 KiB}
            & \multicolumn{2}{c}{64 KiB}
            & \multicolumn{2}{c}{1 MiB}
            & \multicolumn{2}{c}{16 MiB}
            & \multicolumn{2}{c}{64 MiB}
            & \multirow{2}{*}{\mbox{ARE(\%)}} \\
            \cmidrule(lr){2-3}
            \cmidrule(lr){4-5}
            \cmidrule(lr){6-7}
            \cmidrule(lr){8-9}
            \cmidrule(lr){10-11}
            & \commcolhead{AsterMind} & \commcolhead{ns-3}
            & \commcolhead{AsterMind} & \commcolhead{ns-3}
            & \commcolhead{AsterMind} & \commcolhead{ns-3}
            & \commcolhead{AsterMind} & \commcolhead{ns-3}
            & \commcolhead{AsterMind} & \commcolhead{ns-3}
            & \\
            \midrule
            Send/Recv & 3.42 & 4.03 & 4.71 & 6.56 & 25.38 & 28.12
                  & 356.11 & 359.16 & 1414.42 & 1417.46 & 10.78 \\
            AllReduce & 46.86 & 53.32 & 49.12 & 57.76 & 85.30 & 115.96
                      & 664.05 & 706.92 & 2516.10 & 2558.94 & 12.25 \\
            AllGather & 23.95 & 28.19 & 33.00 & 45.89 & 177.69 & 196.82
                      & 2492.75 & 2514.15 & 9900.95 & 9922.24 & 10.78 \\
            AlltoAll  & 53.55 & 56.78 & 56.13 & 59.18 & 97.48 & 88.28
                      & 758.92 & 383.75 & 2875.54 & 1309.77 & 47.71 \\
            \bottomrule
        \end{tabular}%
    }
    \endgroup
\end{table}

\paragraph{Compute model validation.}
We compare the batch completion latency simulated by AsterMind \emph{Fast} mode with that simulated by \emph{Detailed} mode. We evaluate GLM-5.2, DeepSeek-V4 Pro, and Qwen3-32B in the same four-satellite compute cell with TP8~$\times$~PP4 deployment. Each satellite hosts one pipeline stage on eight NVIDIA H200 GPUs in a single scale-up domain. The satellites form a ring with 1~km inter-satellite links configured at 1~Tbps. Both modes use identical batch inputs, rank placements, and link settings. \emph{Detailed} mode uses flow-level bandwidth contention on inter-satellite links, while \emph{Fast} mode only adopts protocol-aware surrogate model with $\Phi_{\text{share}} = 1/0.95$ in Eq.~\eqref{eq:protocol-surrogate} (95\% effective bandwidth). For the Roofline model in Eq.~\eqref{eq:fast-batch-roofline}, \emph{Fast} mode uses 70\% of peak compute throughput and 96.2\% of HBM peak bandwidth.

\begin{figure}[H]
  \centering
  \includegraphics[width=1.0\linewidth]{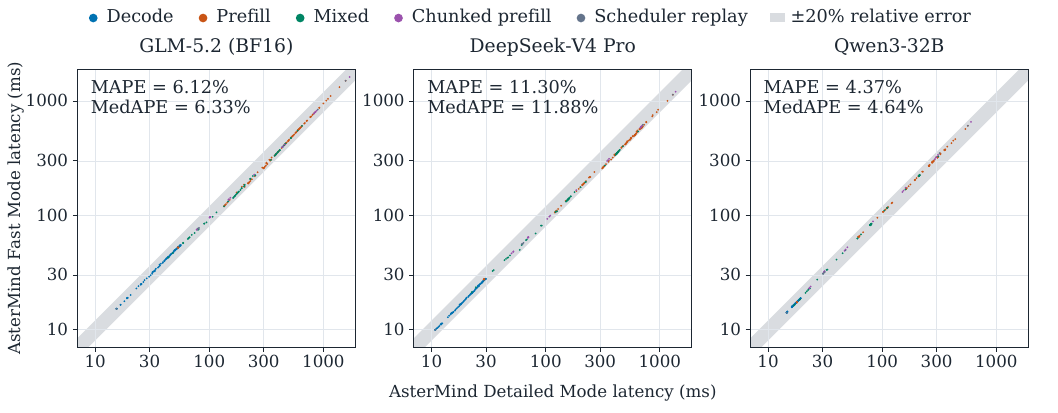}
 \caption{Batch latency comparison on the same TP8~$\times$~PP4 deployment for GLM-5.2 (MoE), DeepSeek-V4 Pro (MoE), and Qwen3-32B (Dense), from left to right. Each model uses the same 275 batches. The gray band marks $\pm 20\%$ error relative to Detailed Mode, and all comparisons fall within this band.}
  \label{fig:Validation_Compute}
\end{figure}

For each model, we evaluate 269 constructed batches and six batches from a scheduler trace. The constructed batches contain 1--64 requests and cover decode-only, prefill-only, mixed prefill and decode, and chunked prefill. We vary context lengths and the proportion of prefill requests in mixed batches. The constructed batches also include prefill requests with cached prompt prefixes to model KV-cache reuse in multi-turn conversations. As shown in Figure~\ref{fig:Validation_Compute}, the mean absolute percentage error relative to \emph{Detailed} mode is 6.12\% for GLM-5.2, 11.30\% for DeepSeek-V4 Pro, and 4.37\% for Qwen3-32B. All 825 comparisons fall within $\pm20\%$ of \emph{Detailed} mode.

\section{Case Studies}
\label{sec:case_studies}
We present three case studies that illustrate how AsterMind can be used to explore orbital LLM serving designs and policies. The first examines how inter-satellite distance and link bandwidth affect inference latency across different parallel deployments within a compute cell. The second examines ground access and compute cell selection under time-varying orbital connectivity. The third compares end-to-end serving latency across runtime policies under the same deployment and workload, using individual requests to explain the differences.

\subsection{Case Study I: Inter-Satellite Links and Inference Latency}
\label{sec:case_study1}

In this section, we use AsterMind to explore how inter-satellite communication affects inference latency and the relative performance of parallelism configurations within a satellite formation.

\textbf{Experimental setup.}
The satellite formation consists of four satellites, each equipped with eight NVIDIA H200 GPUs. The case study is conducted on both DeepSeek-V4-Pro and GLM-5.2. We evaluate 32 requests per DP1 deployment and 128 requests per DP4 deployment (Table~\ref{tab:case-i-deployments}). Each request has 1~K input tokens and 100 output tokens. Chunked prefill is disabled.

\begin{table}[H]
    \centering
    \small
    \caption{Model deployments, cross-satellite traffic, and evaluated requests.}
    \label{tab:case-i-deployments}
    \begin{tabular}{@{}lllr@{}}
        \toprule
        Model & Deployment & Cross-satellite traffic & Evaluated requests \\
        \midrule
        DeepSeek-V4-Pro & TP8 $\times$ PP4 & PP        & 32  \\
        DeepSeek-V4-Pro & TP16 $\times$ PP2 & TP and PP & 32  \\
        DeepSeek-V4-Pro & TP8 / DP4 / EP32 & EP        & 128 \\
        GLM-5.2         & TP8 / DP4 / EP32 & EP        & 128 \\
        \bottomrule
    \end{tabular}
\end{table}

\subsubsection{Link Bandwidth}
\label{sec:case-i-bandwidth}
\begin{figure}[H]
    \centering
    \includegraphics[width=\linewidth]{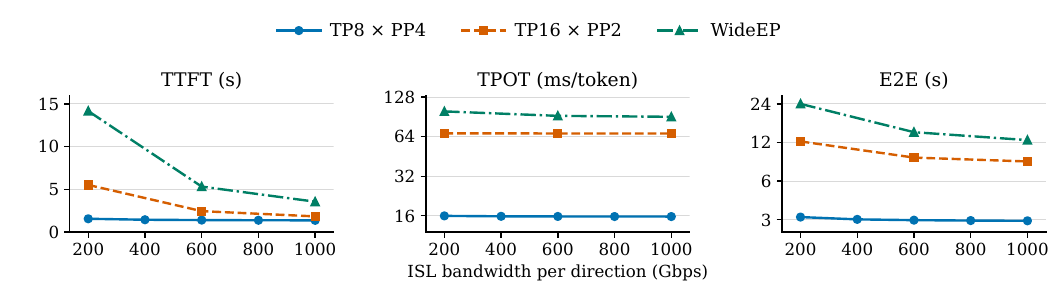}
    \caption{DeepSeek-V4-Pro inference latency as ISL bandwidth varies on a four-satellite ring with 1~km links.}
    \label{fig:case-i-bandwidth}
\end{figure}

For the bandwidth sweep, we use a four-satellite ring with 1~km links and vary per-direction ISL bandwidth from 200 to 1,000~Gbps. As shown in Figure~\ref{fig:case-i-bandwidth}, increasing bandwidth from 200 to 1,000~Gbps reduces TTFT by 11.8\% for TP8 $\times$ PP4, 66.6\% for TP16 $\times$ PP2, and 74.8\% for WideEP. TPOT decreases by less than 0.2~ms/token in both pipeline deployments. For WideEP, it falls from 99.6 to 90.6~ms/token.

\subsubsection{Formation Distance}
\label{sec:case-i-distance}

\begin{figure}[H]
    \centering
    \includegraphics[width=\linewidth]{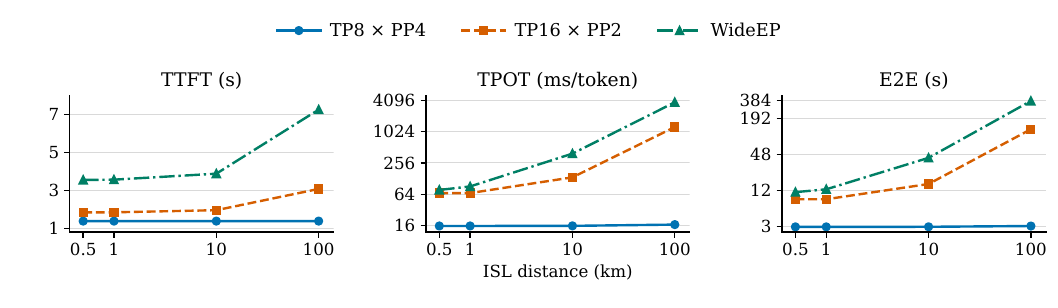}
    \caption{DeepSeek-V4-Pro inference latency as ISL distance varies at 1~Tbps per direction.}
    \label{fig:case-i-distance}
\end{figure}

Fixing per-direction ISL bandwidth at 1~Tbps, we vary the distance between adjacent satellites in the ring formation topology from 0.5 to 100~km. As shown in Figure~\ref{fig:case-i-distance}, increasing the distance from 1 to 100~km raises TPOT by about 1~ms/token for TP8 $\times$ PP4. Over the same range, TPOT rises from 67.6~ms to 1.25~s for TP16 $\times$ PP2 and from 90.6~ms to 3.75~s per token for WideEP deployment. In TP8 $\times$ PP4 deployment, inter-satellite transfers occur at pipeline stage boundaries. In contrast, TP16 $\times$ PP2 and WideEP introduce repeated inter-satellite TP and EP exchanges, making TPOT more sensitive to satellite separation.

\subsubsection{Formation Topology}
\label{sec:case-i-topology}
\begin{figure}[b!]
    \centering
    \includegraphics[width=0.9\linewidth]{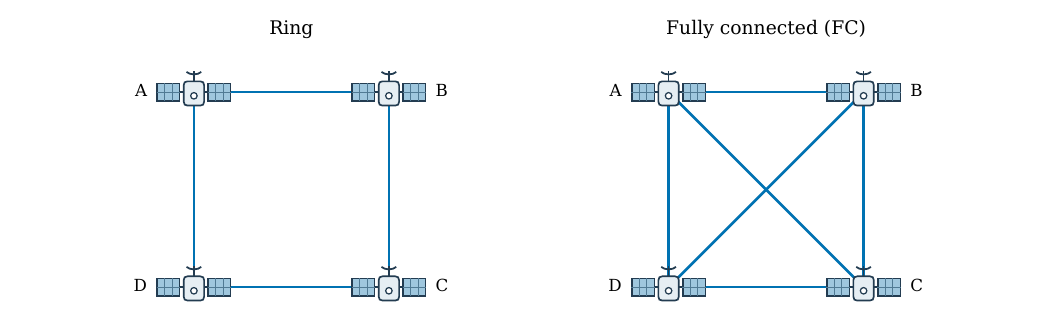}
    \par\smallskip
    \includegraphics[width=\linewidth]{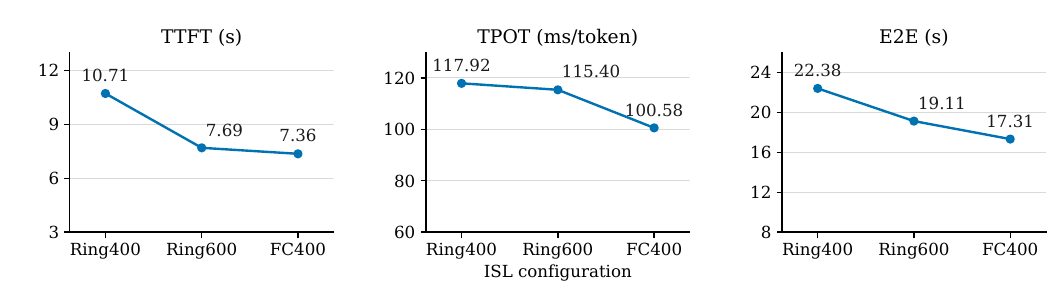}
    \caption{Ring and fully connected ISL topologies (top) and GLM-5.2 WideEP inference latency (bottom). A--D denote the four satellites. Each latency marker shows the median across 128 completed requests.}
    \label{fig:case-i-topology}
\end{figure}
We also examine how formation topology affects inference latency within a compute cell. Fixing the ISL lengths to 1~km, we run GLM-5.2 with WideEP deployment on three network configurations. In Ring400 and Ring600 configurations, each satellite connects to its two neighbors, forming a four-link ring with per-direction link rates of 400 and 600 Gbps, respectively. FC400 adds two direct links on the Ring400, connecting the opposite satellite pairs as shown in Figure~\ref{fig:case-i-topology}.

FC400 and Ring600 both have a summed nominal link rate of 2.4~Tbps per direction. Relative to Ring600, FC400 reduces TTFT by 4.3\% and TPOT by 12.8\%. E2E latency drops from 19.11 to 17.31~s, a 9.4\% reduction. The added direct links shorten the path between opposite satellites from two hops to one. FC400 reduces E2E latency by 22.6\% at the same per-link rate.



\subsection{Case Study II: Ground Access and Compute Cell Selection}
\label{sec:dawn-dusk-session-routing}
In this section, we use AsterMind to examine how ground access and service cell selection affect routing efficiency and effective service capacity as orbital connectivity changes.

\paragraph{Experimental setup.} The compute satellites occupy dawn--dusk Sun-synchronous orbits, crossing the equator near 06:00 and 18:00 local solar time. A near-polar relay constellation connects the ground stations to compute cells throughout the day. All experiments cover a 24-hour horizon and use the same network topologies, sampled at 60~s intervals. Detailed setup is demonstrated in Table~\ref{tab:case-ii-settings}. The link establishment is modeled as described in Section~\ref{sec:Constellation_modeling} and Section~\ref{sec:network_modeling}.

\begin{table}[H]
  \centering
  \caption{Constellations and communication configurations used in Case Study~$\mathrm{II}$.}
  \label{tab:case-ii-constellations}
  \label{tab:case-ii-settings}
  \small
  \setlength{\tabcolsep}{5pt}
  \renewcommand{\arraystretch}{1.10}
  \begin{tabularx}{\linewidth}{@{}lXX@{}}
    \toprule
    \textbf{Constellation} & \textbf{Service cells} & \textbf{Relay satellites} \\
    \midrule
    Planes $\times$ satellites per plane & $4 \times 64$ & $18 \times 36$ \\
    Total satellites & 256 & 648 \\
    Altitude & 600~km & 500~km \\
    Inclination & $97.79^\circ$ & $87.9^\circ$ \\
    Cells per plane & 16 & --- \\
    Satellites per cell & 4 & --- \\
    Along-track spacing within a cell & 1~km & --- \\
    External links per cell & 1 RF, up to 4 optical & --- \\
    \midrule
    \textbf{Link} & \textbf{Data rate} & \textbf{Establishment time} \\
    \midrule
    Ground RF ($25^\circ$ min.\ elevation) & 50~Mb/s up, 100~Mb/s down & 100~ms \\
    Relay--relay optical & 25~Gb/s & 2~s \\
    Relay--cell optical & 10~Gb/s & 2~s \\
    \bottomrule
  \end{tabularx}
\end{table}

\subsubsection{Relay-Only and Hybrid Ground Access}
\label{sec:dawn-dusk-access}
We first compare \emph{relay-only} access with \emph{hybrid} access that permits direct ground-to-cell links. At each time step, we simulate sending one request from each of the four stations (Singapore, Shanghai, Beijing, and Troms\o{}) to the compute cell closest to that station. Under \emph{relay-only} access policy, the station uploads the request to a relay satellite, and the relay network forwards it to the target cell. Under \emph{hybrid} access policy, the station can establish a direct link when the target cell is visible and the direct transfer path is shorter than the relay path.

\begin{figure}[t!]
    \centering
    \begin{subfigure}[b]{0.32\linewidth}
        \centering
        \includegraphics[width=\linewidth]{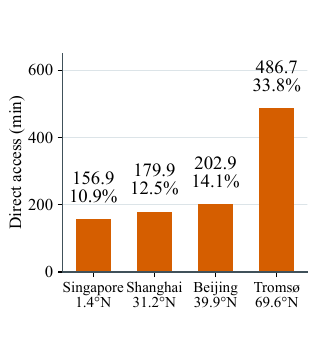}
        \caption{Direct-access availability}
        \label{fig:direct-access-time}
    \end{subfigure}\hfill
    \begin{subfigure}[b]{0.32\linewidth}
        \centering
        \includegraphics[width=\linewidth]{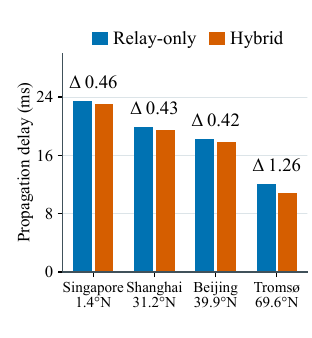}
        \caption{Mean propagation delay}
        \label{fig:mean-nearest-propagation}
    \end{subfigure}\hfill
    \begin{subfigure}[b]{0.32\linewidth}
        \centering
        \includegraphics[width=\linewidth]{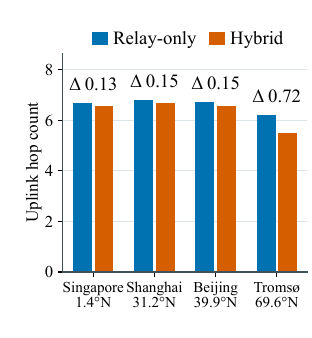}
        \caption{Mean hop count}
        \label{fig:mean-nearest-hops}
    \end{subfigure}
    \caption{Comparison of relay-only and hybrid ground access from four ground stations to the nearest compute cell over a 24-hour horizon. $\Delta$ denotes the reduction achieved by hybrid access relative to relay-only access.}
    \label{fig:dawn-dusk-access-comparison}
\end{figure}

As shown in Figure~\ref{fig:dawn-dusk-access-comparison}(a), Troms\o{} could connect directly to at least one cell for 33.8\% of the day, compared with 10.9\% in Singapore. Under hybrid access policy, only 281 of the 1,440 requests from Troms\o{} (19.5\%) used a direct link to their target cell. Across all requests from Troms\o{}, hybrid access reduces one-way propagation delay by 1.26~ms and hop count by 0.72 relative to relay-only access (Figures~\ref{fig:dawn-dusk-access-comparison}(b) and~(c)). In the other three stations, the corresponding average reductions were below 0.5~ms and 0.16 hops. These evaluation results demonstrate that hybrid access yields only marginal reductions in propagation time and hop count compared to the relay-only access policy.

\subsubsection{Cell Assignment under Concurrent Global Demand}
\label{sec:dawn-dusk-global-demand}

\begin{figure}[!b]
    \centering
    \begin{subfigure}[b]{0.48\linewidth}
        \centering
        \includegraphics[width=\linewidth]{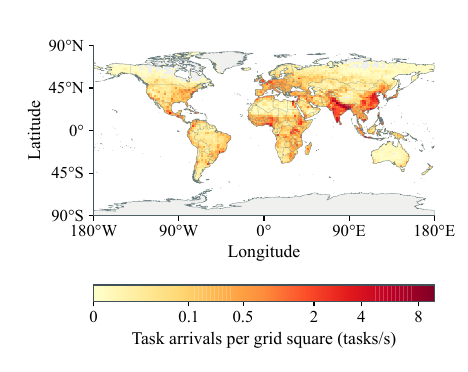}
        \caption{Offered task rate per populated grid square}
        \label{fig:global-demand-map}
    \end{subfigure}\hfill
    \begin{subfigure}[b]{0.48\linewidth}
        \centering
        \includegraphics[width=\linewidth]{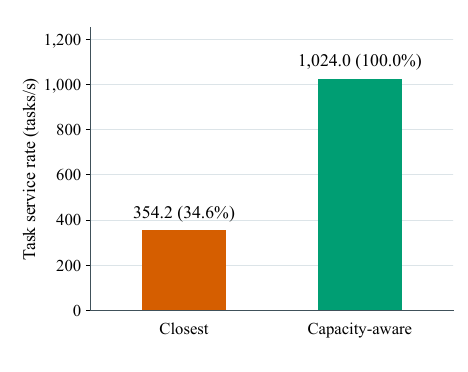}
        \caption{Mean task rate served by 64 cells}
        \label{fig:global-served-rate}
    \end{subfigure}
    \caption{Concurrent global demand distribution and cell assignment performance across 64 cells (total offered load of 1,024~tasks/s). In (b), the percentage above each bar indicates the fraction of the offered load that is served.}
    \label{fig:dawn-dusk-global-demand}
\end{figure}

With concurrent demand, several relays can send traffic to the same cell even when other cells have capacity to spare. To measure how much capacity users can reach, we simulate the workload of 1,024~tasks/s, equal to the combined capacity of 64 cells at 16~tasks/s each. We model independent tasks as steady traffic rates and apply the capacity limits in each one-minute interval. The simulated workload follows the projected residential population in 2025. We aggregate a 30-arcsecond global population raster~\cite{schiavina2023ghspop} onto a $2^\circ \times 2^\circ$ grid and allocate the offered load across the 4,665 populated grid squares in proportion to their populations, as shown in Figure~\ref{fig:dawn-dusk-global-demand}(a). Each square uses its population-weighted center as the ingress location and selects the closest visible relay, retaining it until contact ends.

We evaluate two cell assignment policies, \emph{Closest} and \emph{Capacity-aware}. Under both policies, each relay receives the same load of ground tasks. \emph{Closest} dispatches all tasks directly to the nearest reachable serving cell. By contrast, \emph{Capacity-aware} accounts for tasks already assigned by other relays and selects the nearest reachable serving cell with spare service capacity, assigning tasks up to its remaining limit. 
As shown in Figure~\ref{fig:dawn-dusk-global-demand}(b), under the \emph{Closest} policy, the constellation serves an average of 354.2~tasks/s, accounting for 34.6\% of the offered load. The highest assigned task rate across cells in each minute averaged 221~tasks/s over the day, nearly 14 times the per-cell service limit of 16~tasks/s. By contrast, the \emph{Capacity-aware} policy redistributed excess demand to reachable cells with spare capacity and served the full offered load of 1,024~tasks/s.

\subsection{Case Study III: Request Scheduling for Global LLM Serving}
\label{sec:case_study3}

\begin{table}[!b] \centering \caption{Constellations and communication configurations used in Case Study~$\mathrm{III}$.} \label{tab:case-iii-links} \label{tab:case-iii-replay} \small \setlength{\tabcolsep}{5pt} \renewcommand{\arraystretch}{1.10} \begin{tabularx}{\linewidth}{@{}lXX@{}} \toprule \textbf{Constellation} & \textbf{Service cells} & \textbf{Relay satellites} \\ \midrule Deployment & 10 cells with 4 satellites each & 60 planes with 100 satellites each \\ Total satellites & 40 & 6,000 \\ Altitude & 550~km & 500~km \\ Inclination & Five cells at $53^\circ$ and five in dawn--dusk orbits at about $97.59^\circ$ & $53^\circ$ \\ Topology & Four-satellite ring with 1~km links & Four-neighbor grid \\ Optical terminals & One access satellite per cell, with two ring terminals and two external terminals & Six shared terminals per satellite \\ \midrule \textbf{Link} & \textbf{Data rate} & \textbf{Establishment time} \\ \midrule Ground RF ($25^\circ$ min.\ elevation) & 400~Mb/s up, 1~Gb/s down & 2~s \\ Relay--relay optical & 400~Gb/s per direction & 2~s \\ Relay--cell optical & 400~Gb/s per direction & 2~s \\ Intra-cell optical & 1~Tb/s per direction & 2~s \\ \bottomrule \end{tabularx} \end{table}

As satellites move, the set of compute cells reachable from each ground gateway changes, as do their forwarding delays. For orbital LLM serving, requests from different regions arrive throughout the day and change the queues at those cells. In a multi-turn session, the cell selected for one turn also affects where the next turn can reuse its KV cache. Cell selection therefore affects both the current request and subsequent turns. In this section, we use AsterMind to replay global workload traces and evaluate different request scheduling policies, including \emph{Nearest}, \emph{Session-fixed}, and \emph{Least-loaded}.

\paragraph{Experimental setup.}
We build a 24-hour multi-turn workload from WildChat~\cite{zhao2024wildchat}, which consists of 215,722 requests. During replay, each request arrives at its assigned regional gateway at the time specified in the trace. 
All three policies use the same network topology snapshots, sampled every 60~s over the 24-hour replay. Table~\ref{tab:case-iii-links} lists the constellation and link parameters. Requests are routed along paths that minimize propagation delay.

Each cell serves GLM-5.2 in BF16 on eight H200 GPUs per satellite, with TP8 within each satellite and PP4 across the four satellites. For each GPU, \emph{Fast} mode sets the effective compute throughput and HBM bandwidth to 70\% and 96\% of their peak values. We use vLLM 0.30.0 with continuous batching and chunked prefill, allowing up to 32 concurrent requests per cell and a budget of 16,384 tokens per batch. The cell sends the first token as soon as it is generated and returns the remaining tokens together after decoding. For three scheduling policies, \emph{Nearest} selects the reachable cell whose access satellite is closest to the gateway on each turn. \emph{Session-fixed} applies the same rule on the first turn and retains that assignment for the rest of the session. \emph{Least-loaded} selects the reachable cell with the fewest assigned requests awaiting compute completion.


\subsubsection{End-to-End Serving Performance}

\begin{figure}[!b]
\centering
\captionsetup{skip=3pt}
\includegraphics[width=\linewidth]{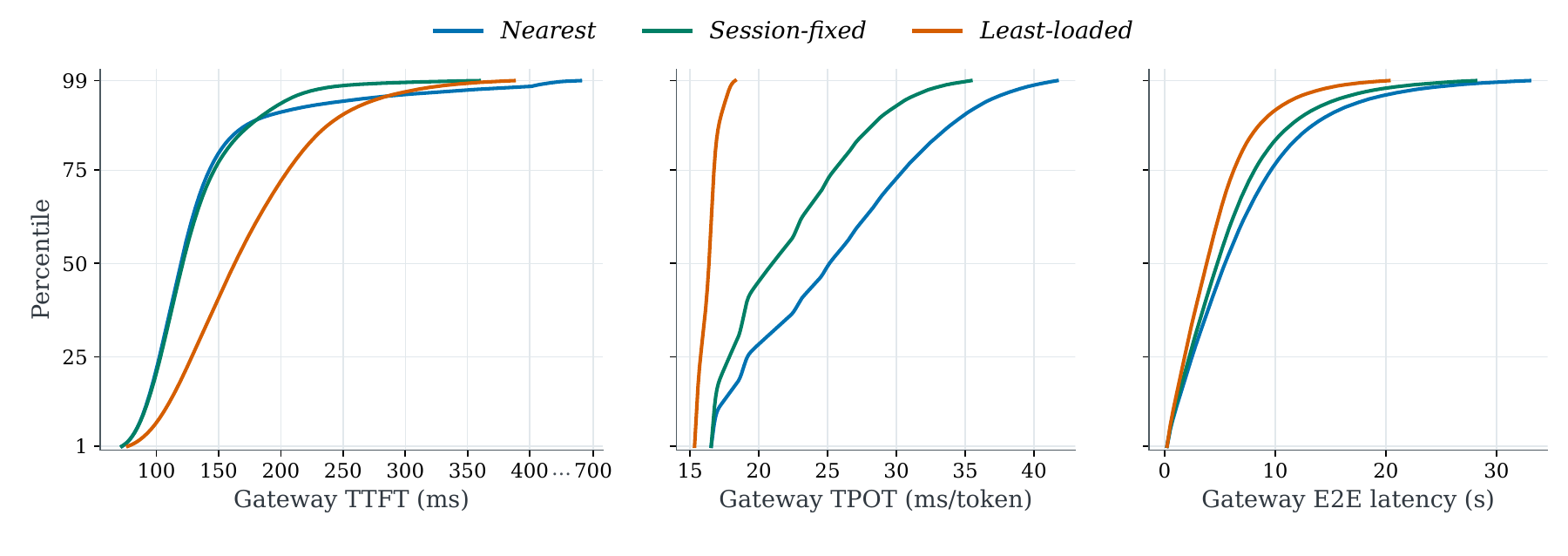}
\caption{Serving performance of different runtime scheduling policies. From left to right, the panels show TTFT, TPOT, and E2E latency distribution.
}
\label{fig:case-iii-performance}
\end{figure}

Table~\ref{tab:case-iii-latency} and Figure~\ref{fig:case-iii-performance} report the serving performance of the requests in one day under all three policies. \emph{Nearest} and \emph{session-fixed} had similar median TTFTs, at 119.5 and 121.0~ms, respectively. After the crossover, \emph{session-fixed} maintains a lower TTFT than \emph{nearest}, where its TTFT was 359.3~ms compared with 639.1~ms at P99. \emph{least-loaded} achieves the lowest TPOT consistently and the lowest end-to-end latency at nearly all the percentiles. Compared with \emph{nearest}, it reduces TPOT by up to 56.2\% and end-to-end latency by up to 38.9\%.

\begin{table}[!b]
  \centering
  \small
  \caption{Latency percentiles over the 24-hour workload replay.}
  \label{tab:case-iii-latency}
  \setlength{\tabcolsep}{3pt}
  \renewcommand{\arraystretch}{1.10}

  \begin{tabular}{@{}lrrr@{\hspace{8pt}}rrr@{\hspace{8pt}}rrr@{}}
    \toprule
\multirow{2}{*}[-2pt]{Policy}
  & \multicolumn{3}{c}{TTFT (ms)}
  & \multicolumn{3}{c}{TPOT (ms/token)}
  & \multicolumn{3}{c}{E2E latency (s)} \\
\cmidrule(lr){2-4}\cmidrule(lr){5-7}\cmidrule(lr){8-10}
  & P50 & P95 & P99
  & P50 & P95 & P99
  & P50 & P95 & P99 \\
\midrule
    \emph{Nearest}
      & \textbf{119.5} & 289.8 & 639.1
      & 24.67 & 37.35 & 41.68
      & 5.44 & 19.72 & 32.98 \\
    \emph{Session-fixed}
      & 121.0 & \textbf{212.3} & \textbf{359.3}
      & 19.94 & 31.03 & 35.41
      & 4.75 & 16.76 & 28.11 \\
    \emph{Least-loaded}
      & 162.8 & 286.9 & 387.4
      & \textbf{16.28} & \textbf{17.70} & \textbf{18.28}
      & \textbf{3.79} & \textbf{12.36} & \textbf{20.28} \\
    \bottomrule
  \end{tabular}
\end{table}

\subsubsection{KV-cache Reuse}
\label{sec:kv-cache_reuse}

\emph{Session-fixed} policy assigns later turns of a session to the same cell, allowing them to reuse the session's KV cache. Across all the later-turn requests, the proportions of requests that reused KV Cache are 98.0\%, 56.1\%, and 7.6\% under \emph{session-fixed}, \emph{nearest} and \emph{least-loaded} policies, respectively. For the 1,178 requests with at least 4,096 tokens of prior context, median TTFT was 182.6~ms under \emph{session-fixed} and 462.1~ms under \emph{least-loaded}. Take one request from Guangzhou as an example. Although \emph{session-fixed} policy took 27.1~ms longer than \emph{least-loaded} to reach the cell, it reused KV-cache for 1,920 tokens. As a result, the gateway received the first token response 26.4~ms earlier. After the first token generation, median TPOT was 16.22~ms/token under \emph{least-loaded} lower than \emph{session-fixed} policy. KV-cache reuse can shorten TTFT by reducing prefill work, while the load at the serving cell also affects how quickly the remaining tokens are generated.

\subsubsection{Uplink Propagation and Cell Load} 
\label{sec:uplink_propagation_and_cell_load}

\begin{figure}[h]
\centering
\includegraphics[width=0.95\linewidth]{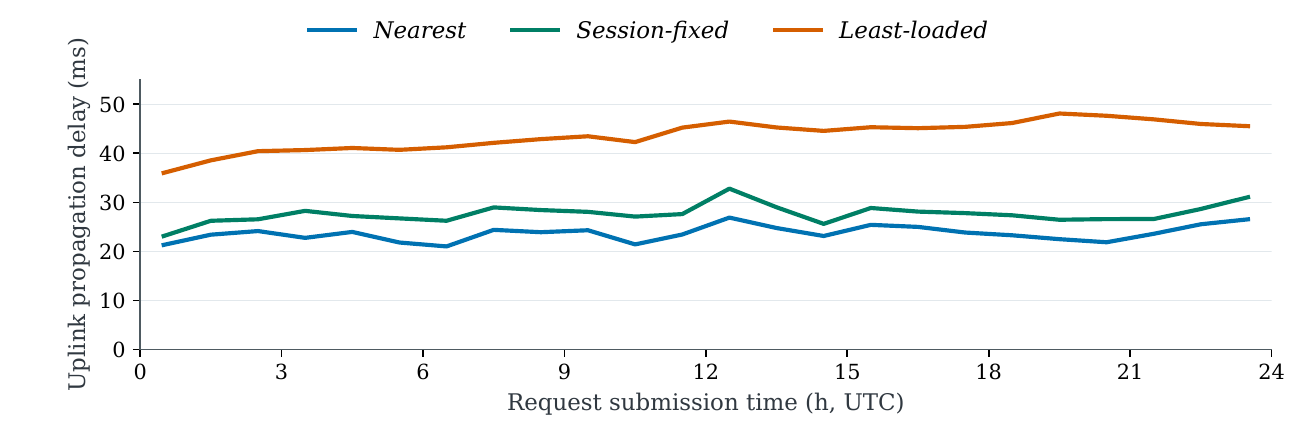}
\caption{One-way uplink propagation delay from active gateways to the cell
selected by each policy. Requests are grouped by submission hour at the gateway, and each point shows the median delay for that hour.}
\label{fig:case-iii-uplink}
\end{figure}

As Figure~\ref{fig:case-iii-uplink} shows, \emph{nearest} had the lowest median uplink propagation delay in every hour of the replay. Across the full replay, mean uplink delay was 24.7~ms under \emph{nearest}, 31.6~ms under \emph{session-fixed}, and 46.3~ms under \emph{least-loaded}. The results help explain why \emph{nearest} policy achieves the lowest median TTFT. 

Serving latency also depends on the load of the selected cell. For one request from Beijing, uplink propagation took 20.8~ms under \emph{nearest} and 35.3~ms under \emph{least-loaded}. The request then waits 29.7~ms for execution under \emph{nearest} and start immediately under \emph{least-loaded}. The queueing delay outweighed the propagation savings, exhibiting TTFTs of 127.1 and 118.5~ms, respectively.


The above results demonstrate that scheduling policies improve different aspects of end-to-end service performance. Shorter forwarding paths help reduce TTFT for typical requests, spreading requests across different cells reduces tail E2E latency, and preserving session locality reduces prefill work for long follow-up requests. AsterMind helps users compare the end-to-end performance of
scheduling policies under specific workloads and constellation configurations, identifying the bottlenecks behind the observed differences.

\section{Conclusion}
\label{sec:conclusion}

We present AsterMind V1, an extensible simulation framework for exploring orbital LLM serving. As the first open-sourced simulation platform for space intelligence computing, AsterMind supports end-to-end simulation across satellite, formation and constellation-level scenarios. It jointly models orbital dynamics, communication and computation on a shared event timeline to capture the interaction between orbital and runtime dynamics. AsterMind supports configurable scenarios and extensible runtime policies, allowing users to compare deployment configurations and scheduling strategies flexibly.

\bibliographystyle{unsrt}
\bibliography{reference}
\end{document}